\documentclass[twocolumn]{openjournal}
\usepackage{amsmath} 

\usepackage{amsmath}

\usepackage{xcolor}
\usepackage{textgreek}
\usepackage[utf8]{inputenc}
\usepackage[english]{babel}

\usepackage{hyperref}
\hypersetup{
    unicode, 
    colorlinks=true,
    linkcolor=linkcolor,
    citecolor=linkcolor,
    filecolor=linkcolor,
    urlcolor=linkcolor,
}
\usepackage{color,colortbl}
\definecolor{linkcolor}{rgb}{0.0,0.3,0.5}
\usepackage{tensind}
\tensordelimiter{?}
\DeclareGraphicsExtensions{.bmp,.png,.jpg,.pdf}
\usepackage{verbatim}
\usepackage[normalem]{ulem}
\usepackage{orcidlink}
\usepackage{soul}
\usepackage{fontawesome}

\graphicspath{{./} {./figs/} }

\newcommand{\ser}{S\'ersic}
\newcommand{\radF}{$\mathcal{F}_r$}
\newcommand{\radFemul}{$ \tilde{ \mathcal{F}_r }$}

\begin{document}
\title{Using Symbolic Regression to Emulate the Radial Fourier Transform of the S\'ersic profile for Fast, Accurate and Differentiable Galaxy Profile Fitting\vspace{-.5in}}
\author{Tim B. Miller\orcidlink{0000-0001-8367-6265}}
\affiliation{Center for Interdisciplinary Exploration and Research in Astrophysics (CIERA), Northwestern University, 1800 Sherman Ave, Evanston, IL 60201, USA}

\author{Imad Pasha\orcidlink{0000-0002-7075-9931}}
\affiliation{Astronomy Department, Yale University, 219 Prospect St,
New Haven, CT 06511, USA}
\affiliation{Dragonfly Focused Research Organization, 150 Washington Avenue, Suite 201, Santa Fe, NM 87501, USA}

\shorttitle{Emulating the Radial Fourier Transform of the S\'ersic Profile}
\shortauthors{Miller \& Pasha}
\begin{abstract}
Galaxy profile fitting is a ubiquitous technique that provides the backbone for photometric and morphological measurements in modern extragalactic surveys. A recent innovation in profile fitting algorithms is to render, or create, the model profile in Fourier space, which aims to provide faster and more accurate results. However, the most common parameterization, the \ser~profile, has no closed form Fourier transform, requiring the use of computationally expensive approximations. In this paper our goal is to develop an emulator to mimic the radial Fourier transform of the \ser~profile, for use in profile fitting. We first numerically compute the radial Fourier transform and demonstrate that it varies smoothly as a function of the \ser~index and $k$, the spatial frequency coordinate. Using this set of numerically calculated transforms as a training set, we use symbolic regression to discover an equation which approximates its behavior. This ensures the emulator will be based on computationally efficient and differentiable building blocks. We implement this novel rendering method in the \texttt{pysersic} profile fitter, and ensure it is accurate by conducting both injection-recovery tests using model galaxy profiles and applying multiple rendering methods to a real sample of galaxies in HSC-SSP imaging. Crucially, the Fourier emulator rendering technique enables measurements of morphological parameters of galaxies 2.5 times faster than standard methods with minimal loss in accuracy. This increased performance while maintaining accuracy is a step that ensures these tools can continue to scale with the ever-increasing flow of incoming data. 
\end{abstract}

\maketitle

\section{Introduction}
\label{sec:intro}

Galaxy profile fitting describes a forward modeling approach to measuring the structural parameters and photometry of galaxies from their images. In many cases the galaxy is assumed to be axisymmetric, with a parametric form of the radial surface brightness profile. The best fit parameters are used to investigate the shape of the surface brightness profile, size, and total flux of the galaxy. By directly modeling the image, one is able to account for the effect of the point spread function, or PSF. Profile modeling has thus become a ubiquitous tool for modern extragalactic astronomy for investigating the morphological properties of galaxies~\citep[e.g.][]{Lange:2015,Mowla:2019,Kawinwanichakij:2021, Ward:2024, Allen:2024}. For large surveys it is used to obtain accurate photometry of extended sources~\citep{Bernardi2013}. Such modeling is also necessary to accurately model sources in crowded fields and across instruments with vastly different depths and spatial resolutions~\citep{Dey2019,Weaver2022}.

In many of these applications galaxy surface-brightness profiles are parameterized using a \ser~profile~\citep{Sersic1963,Sersic:1968}. This empirical description parameterizes the radial surface brightness profile using the functional form shown below:
\begin{equation}
    I(R) \propto F_{\rm total} \exp\left[ -\left( \frac{R}{R_{\rm eff}} \right)^{1/n}  \right]
\end{equation}
Here, $F_{\rm total}$ is the total flux of the galaxy, $n$ is the \ser~index and describes the shape of the profile, and $R_{\rm eff}$ is the effective --- or half-light --- radius, which describes the spatial scale of the galaxy. These three parameters, along with the axis ratio, position angle and central position, are used to parameterize the light profile of galaxies. The introduction of the index, $n$, consolidated the fixed index profiles such as the de Vaucouleurs~\citep{devaucouleurs1948} and exponential into a single generalized parametrization. Given this additional freedom the \ser~profile is able to provide a reasonable approximation to the surface brightness profiles of almost all galaxies. It is also possible to create non-axisymmetric models from this base profile by adding Fourier modes or warps~\citep{Peng:2002,Stone:2023}. There are numerous software packages and methods that perform \ser~profile fitting, each with their own aims and strengths. Some of the most popular examples are: \texttt{galfit}~\citep{Peng:2002}, \texttt{imfit}~\citep{Erwin:2015}, \texttt{profit}~\citep{Robotham:2017}, \texttt{astrophot}~\citep{Stone:2023}, \texttt{pysersic}~\citep{Pasha:2023} and \texttt{forcepho}~\citep{Baldwin2024}.

While the \ser~profile's ability to characterize real galaxies has made it the most popular parametrization in the literature, the use of \ser~functions in profile fitting is accompanied by several numerical challenges. The first concern is how to accurately render model images, i.e., how to assign model flux values at each pixel given a set of \ser~ parameters. This is a particular problem for galaxies with large indices, $n\gtrsim 2$. Near the center of these galaxies, the slope of the surface brightness profile becomes very steep, leading to drastic changes on scales smaller than that of a pixel. Naive implementations which assign fluxes based on the midpoint of the pixel will fail for these profiles, as they implicitly assume the flux at the center of the pixel provides a reasonable approximation for the average of that pixel --- and for rapidly changing profiles this assumption no longer holds. The most common method to render the centers of profiles accurately is to ``oversample.'' By evaluating the \ser~model at a series of sub-pixel locations and averaging them to produce the flux estimate for a given pixel one obtains accurate model images for steep profiles. Oversampling comes at a computational cost as many more calculations need to be performed per pixel rendered, though it can be performed recursively by checking for convergence to better focus computational resources where necessary~\citep{Robotham:2017, Stone:2023}.

A promising alternative method is to render the galaxy in Fourier space; with this method, similar accuracy in the central region can be achieved without the need for oversampling~\citep{Rowe:2015, Lang:2020}. This provides a more natural way to render the galaxy ``above the atmosphere,'' i.e., the intrinsic galaxy image before including the effects of the PSF and pixelization. Furthermore, a Fourier transform is required to convolve a pixel-rendered image with a PSF regardless; rendering directly in Fourier space negates the need for this initial computation, improving performance. Unfortunately, despite the theoretical benefits of a Fourier representation, the \ser~function has no general closed-form Fourier transform\footnote{At fixed indices such as $n=0.5$, representing a Gaussian, and $n=1$ representing an exponential the Fourier transforms are known analytically}.

A popular workaround, originally suggested in \citet{Hogg2013}, is to approximate the \ser~profile using a mixture of Gaussians model. Gaussians are particularly nice functions to work with, especially when using Fourier transforms, as the transform of a Gaussian is also Gaussian. Gaussian decompositions of \ser~profiles can either be pre-computed and stored in a lookup table~\citep{Rowe:2015, Lang:2016} or calculated on the fly, such as using the algorithm presented in \citet{Shajib:2019}. Additionally, if the PSF is also modeled using Gaussians, then the convolution can be performed analytically. The Mixture of Gaussians representation has been implemented in packages for \ser~profile fitting such as \texttt{tractor}~\citep{Lang:2016}, \texttt{pysersic}~\citep{Pasha:2023}  and \texttt{forcepho}~\citep{Baldwin2024}. Unfortunately this solution is not a panacea --- in order to accurately reproduce the \ser~profiles, a large number of Gaussians ($\gtrsim10$) must be employed. Each Gaussian is rendered independently, thus (again) increasing the computational cost. 

A wholly different approach was put forth in \citet{Spergel2010}. The authors propose a novel functional form of the radial profile, defined in terms of Bessel functions, that loosely mimics the \ser~profile. Their key insight was to specifically design a functional form such that the Fourier transform is convenient to work with. While some studies have utilized this profile~\citep[e.g.][]{Li:2023,Tan2024}, given the decades of inertia of use of the \ser~profile, there is yet to be widespread adoption. 

In this paper, we tackle this longstanding problem using a novel approach. Our first insight is recognizing that although the Fourier transform of the \ser~profile is intractable to calculate analytically, it is possible to calculate numerically. Thus one can easily build up a library of curves that could be used to train an emulator to mimic its behavior. Our second insight is to employ symbolic regression to emulate the Fourier transforms of the \ser~profile. Symbolic regression is a supervised learning task which aims to search a wide space of functional forms to find the best fit to a given set of data~\citep{Angelis2023}. This is usually achieved using a genetic programming approach which begins with a set of simple expressions and combines them to generate more complex equations. For this study we will use symbolic regression to mimic the Fourier transform of the \ser~profile to build an emulator based on simple, computationally efficient and differentiable building blocks. This last point is particularly salient given the rise of gradient based optimization and sampling techniques being implemented for profile fitting~\citep{Stone:2023, Pasha:2023,chen2024}.

The paper is organized as follows: First we calculate the radially symmetric Fourier transform numerically in Section~\ref{sec:num_transform}. We then emulate these curves utilizing the \texttt{pysr} package~\citep{Cranmer2023} to perform symbolic regression in Section~\ref{sec:sr}. In Section~\ref{sec:fitting} we implement a rendering method based on this emulator in the \texttt{pysersic} package to ensure its accuracy for profile fitting. We summarize our approach and discuss caveats and future directions in Section~\ref{sec:summ}. 

\section{The Radially Symmetric Fourier Transform of the \ser~Profile}
\label{sec:num_transform}

\begin{figure*}
    \centering
    \includegraphics[width=0.99\textwidth]{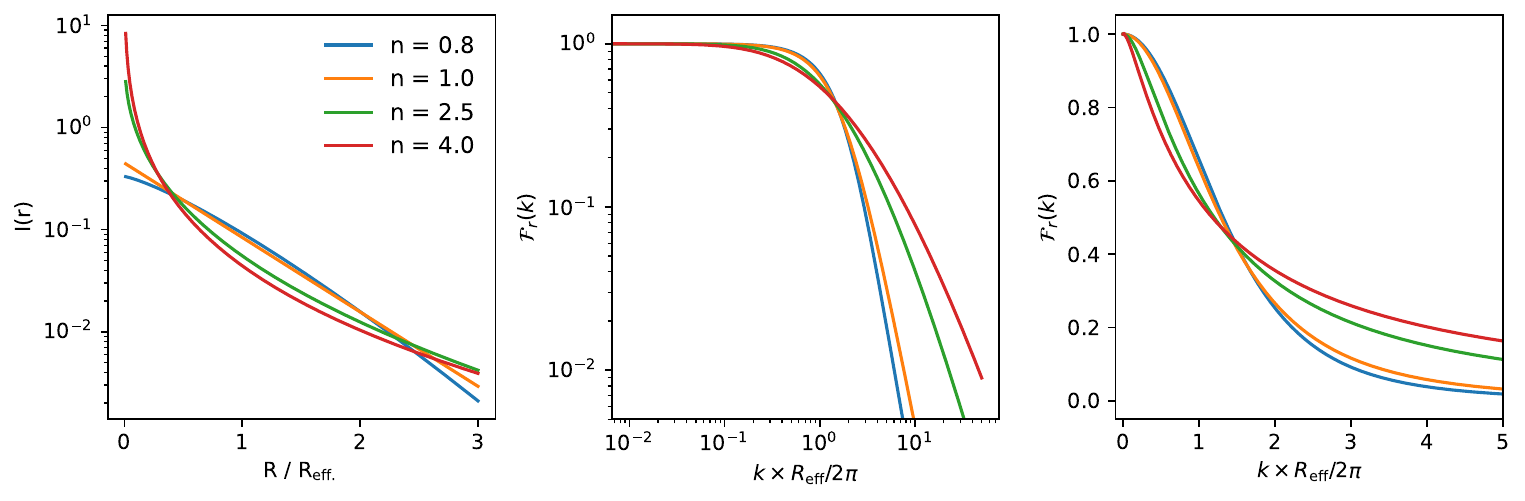}
    \caption{Showcasing the \ser~profile and its associated radial Fourier Transform, also known as the Hankel transform. We show a set of four indices which are all normalized such that the total flux and effective radius are unity. The surface brightness profile is shown (\textit{left}) along with two different views of the radial Fourier transform; log-log (\textit{middle}) and linear (\textit{right}). We find the transformed \ser~profile is bounded between 0 and 1 and asymptotes to 0 at large values of $k$ and 1 as $k\rightarrow0$. Additionally we find smooth variation with $n$.}
    \label{fig:sersic_hankel_showcase}
\end{figure*}

In profile fitting we are interested in the radially symmetric Fourier transform. In almost all applications the galaxy is assumed to be axi-symmetric, with only a radial dependence. To use the Fourier transform we must first switch to using polar coordinates and integrate out the dependence on $\theta$. The radially symmetric Fourier transform can be written as:
\begin{equation}
    \mathcal{F}_r(k) = 2\pi\int_0^\infty I(R) J_0(kR)R dR 
    \label{eqn:hankel}
\end{equation}
where $k$ is the spatial frequency coordinate, and $J_0$ is the Bessel function of the first kind. In two dimensions this is equivalent to another closely related integral transform, the Hankel transform. 

We use the python package \texttt{hankel}~\citep{Murray:2019}, to numerically calculate the radial Fourier transform. By using the numerical integration method for Bessel functions derived in \citet{Ogata:2005}, this package is able to quickly and accurately calculate the Hankel transform for any given function. For all of the calculations performed in this paper, we use a \ser~profile that is normalized with a total luminosity and effective radius equal to unity. For all calculations performed in this study using \texttt{hankel}, the quadrature bin width, $h$, which controls the accuracy of the integrations is set to $10^{-4}$. Correspondingly, the number of quadrature bins, $N$, we set to $\pi/h$ as suggested in the documentation. This level of accuracy is likely unnecessary, as the reported uncertainties are often lower than machine precision, $<10^{-16}$, however all of the transforms needed for this study are able to be calculated within a few minutes.

Examples of the radial Fourier transform of four example \ser~profiles with a range of indices are displayed in Figure \ref{fig:sersic_hankel_showcase}. We highlight \radF~(k) for these four \ser~profiles (left) using two different views of the same data; a log-log (center) and linear (right) plot. We observe that \radF~changes smoothly with both, $k$ and $n$, which is encouraging for the prospects of emulating its behavior. There is also important limiting behavior to consider when approximating \radF. Firstly it is bounded by $[0,1]$. \radF$(k,n)$ tends towards $0$ as $k$ increases and it tends towards 1 as $k$ approaches 0 for all values of $n$. In order to accurately emulate \radF of the \ser~profile it is crucial to match these bounds and limiting behaviors.

\section{Finding an Approximation to \radF~$(k)$ using Symbolic Regression}
\label{sec:sr}
\subsection{Fitting Procedure}
\label{sec:sr_fitting}

\begin{figure*}
    \centering
    \includegraphics[width=0.98\textwidth]{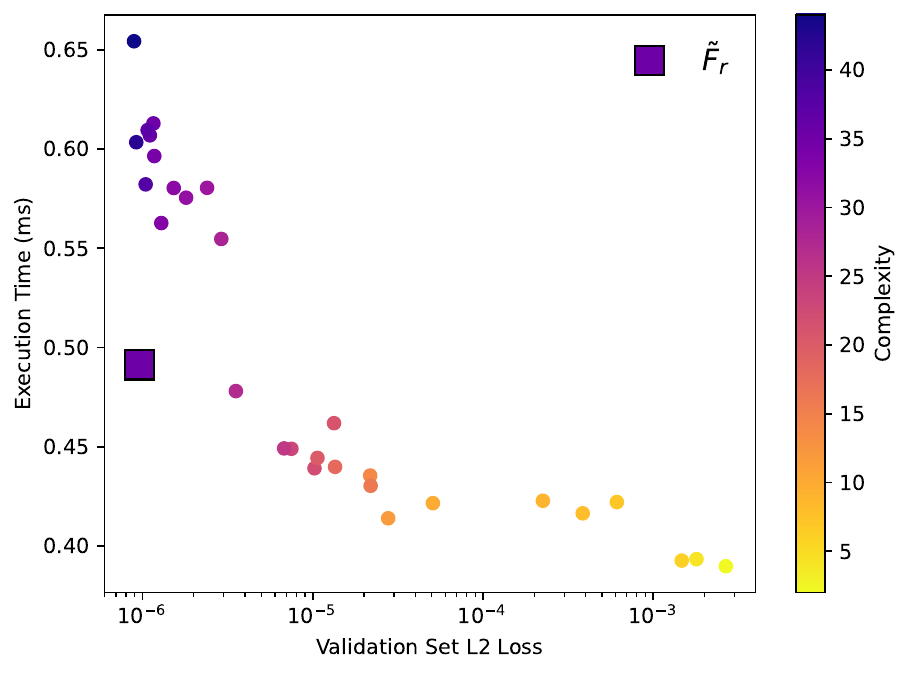}
    \caption{Showcasing the results of the symbolic regression fitting process comparing execution time and accuracy. The execution time is measured for rendering a model image, calculating a $\chi^2$ log-likelihood and calculating the gradients of the input parameters. The accuracy is calculated using the L2 loss, or mean-squared error, of a validation set of $k,n$-\radF\ pairs. Each point is colored by its complexity. We find a negative correlation between execution time and accuracy; more complex equations tend to be more accurate but take longer to compute.  However above a complexity of 30 there is a plateau in accuracy. In order to balance computational efficiency and accuracy we select an equation near the bottom of this plateau, denoted by $\tilde{\mathcal{F}}_r$ and shown in Eqn.~\ref{eqn:emulator}. For comparison we note that the default \texttt{hybrid} renderer in \texttt{pysersic} has an execution time of 1.66 ms, more than twice as slow as the most complex equations presented here and three times slower than our selected equation.}
    \label{fig:loss_time}
\end{figure*}

The goal of symbolic regression is to perform a search of the space of analytical expressions for the expression that best matches a given set of input data~\citep{Angelis2023}. The search begins with a set of simple functions and combines them to form more complex and expressive functions. The equations are typically evolved using a ``genetic'' algorithm wherein the function is continually mutated to search for the equations which best match the data. This is a multi-optimization problem where the goal is to find the best fit to the data while minimizing the complexity of the resulting expression. In this study we perform symbolic regression using the python package \texttt{pysr}~\citep{Cranmer2023}. This package implements an evolutionary algorithm which simultaneously evolves multiple populations of equations at varying complexities. At each step the equations are mutated, i.e. operators are added, removed, or swapped or constants are optimized in order to search for the best representation of the data. The result is a pool of best fitting equations at varying levels of complexity.

Our goal is to find an approximation, \radFemul~$(k,n)$, that mimics the behavior radial Fourier transform of the \ser~profile, \radF~$(k,n)$. To facilitate an easier search and ensure the proper limiting behavior observed in Fig.~\ref{fig:sersic_hankel_showcase}, we will impose the following structure on the resulting equation:

\begin{equation}
    \label{eqn:template}
    \tilde{\mathcal{F}_r}(k,n) = \frac{1}{1 + e^{G(k,n)} }
\end{equation}

where $G(k,n)$ is an arbitrary function that will be searched for using symbolic regression. This constraint is imposed using the template functionality in \texttt{pysr}\footnote{Introduced in version 1.0}. This form is a Sigmoid function with the opposite sign in the exponential to match the limits of the limits of  \radF and enforces the boundedness observed in Fig.~\ref{fig:sersic_hankel_showcase}. Initially in our testing, we did not enforce any structure but found many of the resulting equations eventually converged to a similar form as presented in Eqn.~\ref{eqn:template}. Enforcing this structure a-priori allows for more efficient convergence that results in more accurate final equations. Another option to enforce this boundedness would instead forgo using the template and perform the inverse sigmoid transform before fitting, i.e. performing symbolic regression on $\log\left( \frac{1}{1-\mathcal{F}_r}\right)$. However this, in essence, optimizes for fractional uncertainty on \radF, and places too much emphasis on the high and low values of $k$. Once the sigmoid function has been applied, minor differences in this regime do not alter the behavior of the resulting equation much. Using the template approach allows us to optimize for absolute uncertainty on \radF~rather than fractional uncertainty.

We construct our training set by numerically randomly sampling 10,000 values of $k,n$ and using the \texttt{hankel} python package to calculate \radF~$(k,n)$. In this calculation, the flux and $R_{\rm eff}$ of the profile are set to unity. The resulting equation can be trivially rescaled to any value and this normalization allows us to focus the variation in \radFemul~$(k,n)$ with $k$ and $n$. For $n$ we randomly sample values between 0.4 and 6.5, using a uniform distribution. Our target range for $n$ is $0.5-6$, encompassing the majority of real galaxies ~\citep{Lange:2015,Mowla:2019} and extend the window slightly for the training set to limit the decrease in accuracy near the edges target range. We sample the set of $k$ values from an exponential distribution with a rate of $1/15$. We find that this roughly mimics the distribution of $k$ values when a 200 pixel by 200 pixel image is Fourier transformed. This is around the typical size of cutout used for galaxy image modeling, which is often between 100-200 pixels.~\citep{VanderWel2012,Mowla:2019}.

We allow the following basic binary operators: $+,-,*,$, and the following unary operators: $\exp,\log,{\rm square}, {\rm sqrt}$, inverse and cube. Setting the inverse as a unary operator, instead of the binary division operator, allows us to better control the constraints to avoid numerical issues. We set up a series of constraints on these operators that attempt to avoid circularly nested functions, i.e. a sqrt directly inside a square, or a log directly inside an exponential. Additionally we do not allow any $+$ or $-$ operators inside the inverse to avoid division by 0. In our testing we found this was necessary to ensure numerically stable equations and did not meaningfully alter the final accuracy of the equations.

We use a L2, or mean square error, loss function and a maximum equation complexity of 50. Complexity is measured as the total number of components in an equation; each constant, operator and variable contributes 1 complexity. We use the batching procedure within \texttt{pysr} with a batch size of 500. In our testing we find this was necessary to explore more equations more efficiently. We use 150 population and 2,000 cycles per iteration while setting \texttt{weight\_optimize} parameter to $10^{-3}$ as suggested in the documentation to ensure the constants are optimized. We set the Parsimony and adaptive parsimony scaling to $10^{-8}$ and 500 respectively. These parameters control how much to punish high complexity solutions and we use a lower scaling value to encourage exploration of high complexity, but more accurate, equations. We perform the symbolic search for $2.5 \times 10^6$ iterations, with the results summarized in Figure~\ref{fig:loss_time}.

\subsection{Selecting an Equation}
For many applications, the goal of symbolic regression is to find a simple or interpretable equation. However, our aim is to find an \textit{accurate} and \textit{computationally efficient} equation, with little emphasis on readability or simplicity. To understand the properties of the final population of equations in this context, Figure ~\ref{fig:loss_time} compares the computational efficiency against the accuracy of the final pool of equations. For accuracy we use the L2, or mean squared error, loss for a validation set of 1,000 $k,n$-\radF\ pairs that are drawn from the same distribution as is used for training. For computational efficiency we measure the execution time to render an 150 x 150 pixel image for a given set of parameters and calculate the $\chi^2$ log likelihood, assuming no covariance between pixels, with respect to a reference image. This is implemented in \texttt{jax}, same as \texttt{pysersic}, which allows for auto-differentiation capabilities. In this timing we also include the calculation of gradients of the log likelihood. This setup attempts to mimic  the calculations used for sampling and optimization when these emulators are implemented into \texttt{pysersic}. This timing is repeated 500 times for each equation and we report the median. For comparison we repeat the timing procedure for the \texttt{hybrid} renderer, which serves as the default in \texttt{pysersic} and find an average execution time of 1.66 ms.

\begin{figure*}
    \centering
    \includegraphics[width=0.95\textwidth]{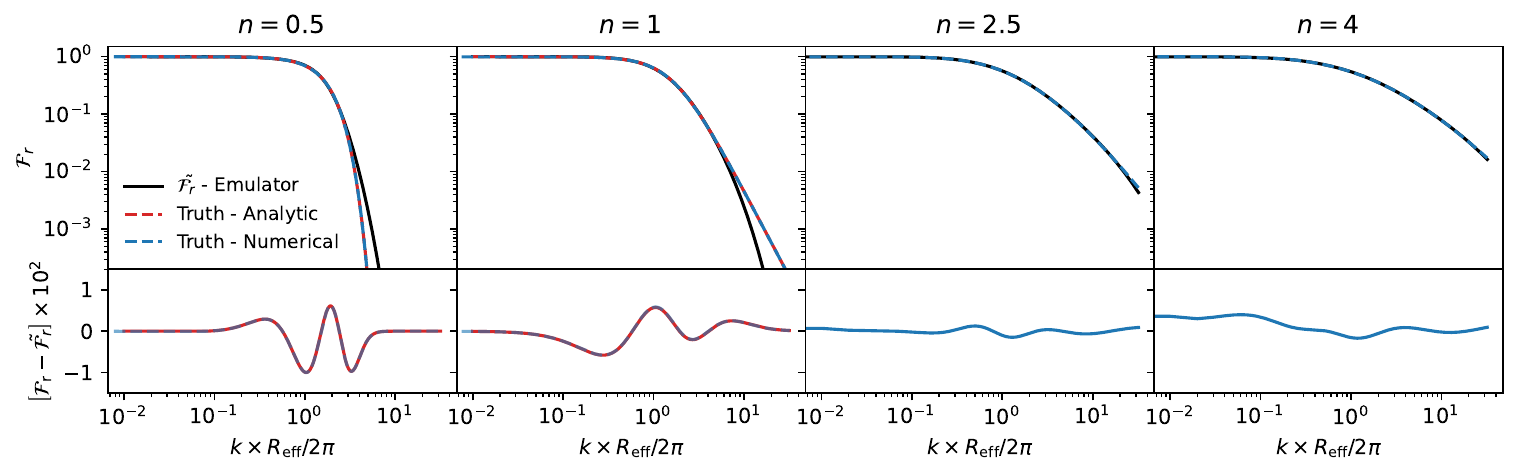}
    \includegraphics[width=0.95\textwidth]{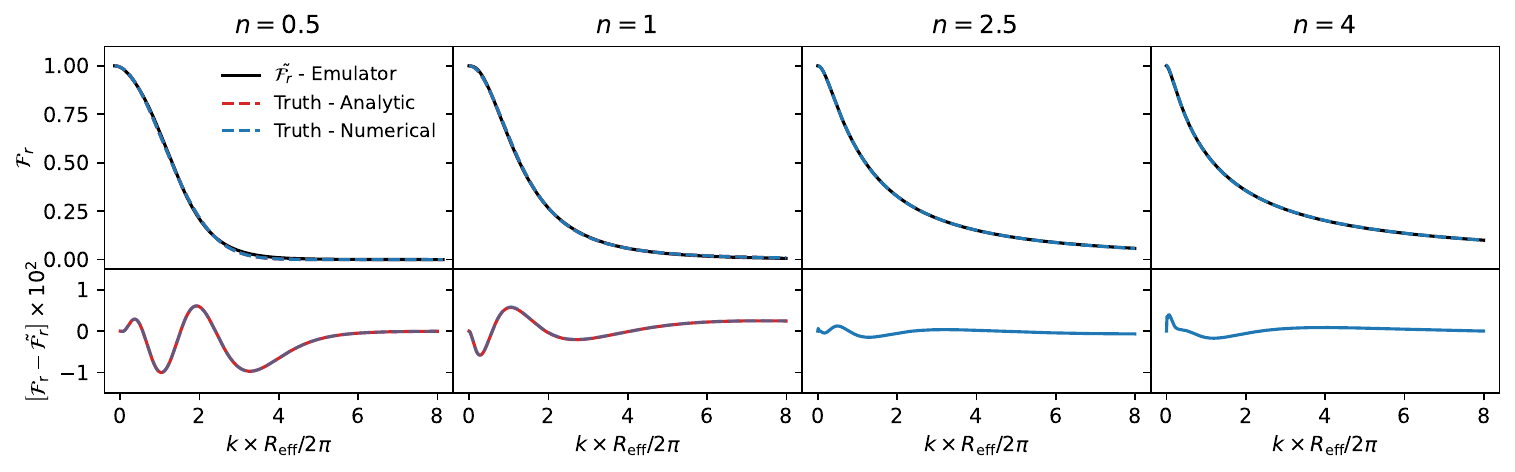}
    \caption{Visualization of the chosen equation compared to analytically or numerically calculated radial Fourier Transforms. We show four profiles, $n=0.5,1,2.5,4$, across two panels which show the profiles with logarithmic (top) and linear (bottom) scalings. In each panel we also show the linear residuals, the same for both panels but with different scaling of the $k$ axis. For $n=0.5$ and $n=1$, we show the analytically and numerically calculated Hankel transforms. For $n=2.5$ and $n=4$ the analytic forms are not known so we rely only on the numerically calculated transforms as the truth.}
    \label{fig:eqn_resid}
\end{figure*}

\begin{figure*}
    \centering
    \includegraphics[width = 0.99\textwidth]{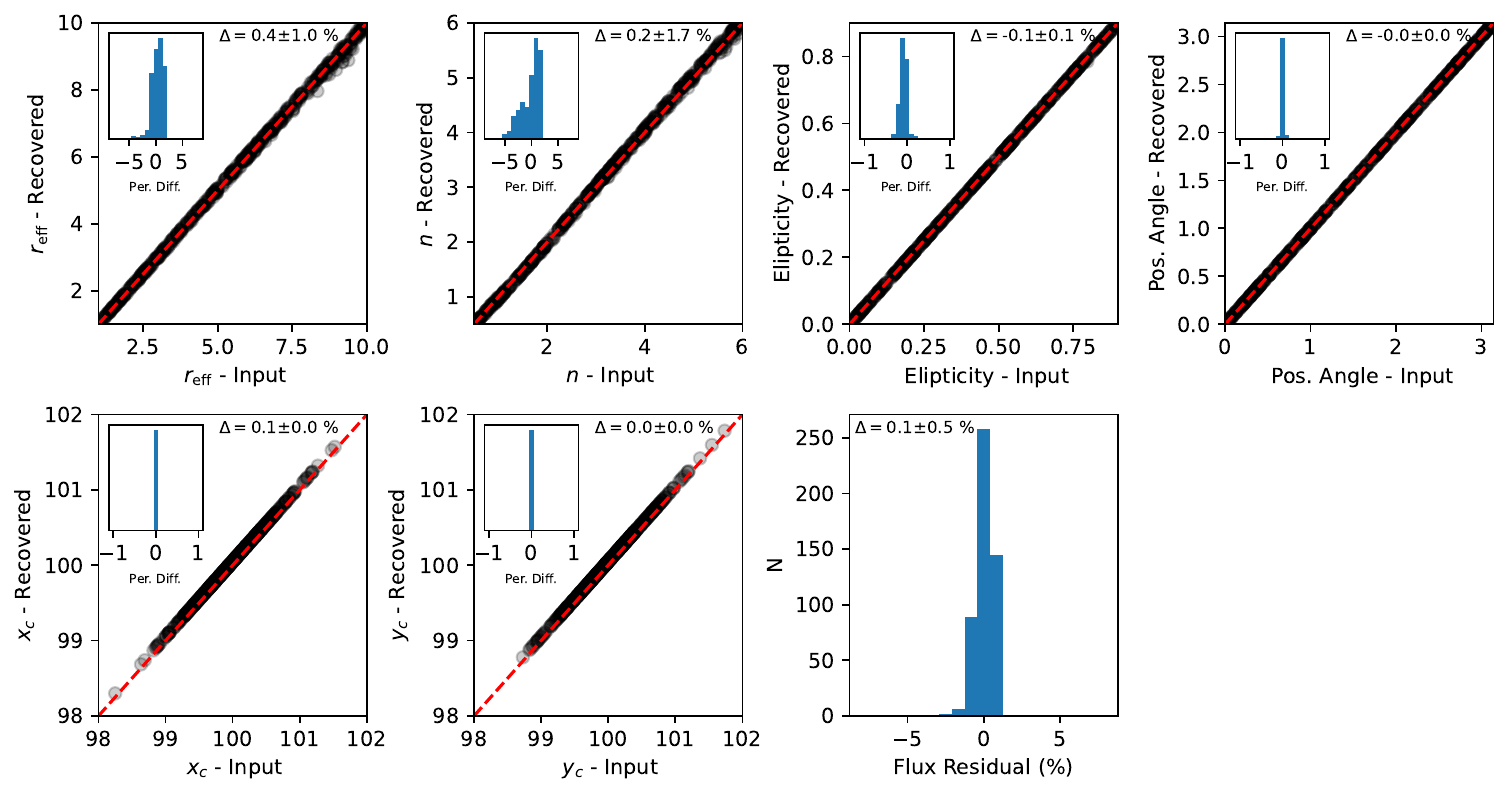}
    \caption{Injection recovery tests to test if our chosen \ser~Fourier Emulator, Eqn.~\ref{eqn:emulator} is accurate for profile fitting. We render a series of profiles using a heavily over-sampled pixel space renderer as the `truth` and fit each using a renderer based on the Fourier Emulator. The input values are compared to those recovered for all of the parameters of the profile. For the flux, since all of the input values are 100, we opt to show a histogram of the residuals. In general we find excellent agreement. For the position, ellipticity and positional angle we find recoveries that are accurate to a fraction of a percent. For $r_{\rm eff}$, the index $n$ and flux the average residual $<0.5\%$ and the scatter $<2\%$, calculated using the biweight location and scale respectively~\citep{Beers1990}, for all three parameters. These results indicate that the rendering method utilizing the \ser~Fourier emulator can be used for profile fitting with minimal loss in accuracy.}
    \label{fig:inj_recov}
\end{figure*}

In general we find there is an inverse correlation between accuracy and computational time. The more complex equations are generally more accurate but require more calculations and therefore have longer execution times. However even these most complex equations still have an execution time over twice as fast as the \texttt{hybrid} renderer. We find there is a plateau in this relationship. Above a complexity of about 30 the execution time continues to increase without any significant gains in the accuracy. In order to maximize both computational efficiency and accuracy we aim to select a point near the beginning of this plateau. Specifically we select the equation with the lowest execution time that has L2 loss $<2 \times 10^{-6}$. This yields the equation with a complexity of 39, which is displayed below, in Equation.~\ref{eqn:emulator}.

\begin{align}
    H(k,n) &= a_0 \sqrt{n + a_1 \left(k - a_2\right) e^{e^{\sqrt{n} - n^{3}}}} \\
    J(k,n) &=  e^{\left(a_3 k  - e^{n - n^2}\right) }\\
    G(k,n) &= \frac{1}{n} \left(\left[ H(k,n) + J(k,n) \right] \left(\log{\left(k \right)} -a_4\right) - a_5 \right)
    \label{eqn:emulator}
\end{align}

We note there is nothing special about the parts of the equation denoted by $H$ and $J$, we have simply separated them out for clarity in presentation. The constants, $a_i$, in Eqn.~\ref{eqn:emulator} are given by:
\begin{equation}
\begin{split}
    a_0 &= 2.245374\\
    a_1 &=  0.029371526\\
    a_2 &= 2.1431181\\
    a_3 &= -3.7275262\\
    a_4 &=  0.091609545\\
    a_5 &=  0.32785136\\
    \label{eqn:emulator_const}
\end{split}
\end{equation}

This equation is an outlier in the plane of execution time and accuracy that makes it particularly suited for our use case. Compared to equations of similar complexity in the final pool there are very few instances of the $k$ evaluation within exponential or square root evaluations. When rendering a profile the number of values of $k$ that need to be calculated scales with the number of pixels and these are computationally expensive operations. Given there is only one value of $n$ to evaluate for each profile the prevalence of this variable has less of an effect on the overall execution time. For the remainder of this paper we will use this equation used to emulate the radial Fourier transform of the \ser~profile and denote it \radFemul.

Our chosen equation is visualized and compared to the true radial Fourier transform in Figure~\ref{fig:eqn_resid}. The emulated equation is shown at four different indices, $n= 0.5, 1, 2.5$ and $4$, alongside the residuals when compared to the true value. For $n=0.5,1$,  we include the analytically calculated radial Fourier transform alongside the numerically calculated one. The selected equation provides a good match to all the profiles across the entire $k$ range shown, with residuals constrained to $<10^{-2}$. The scatter accuracy does depend on $n$, with the values of $n<1$ showcasing larger residuals. The emulated equation does diverge from the truth at high $k$ for these profiles as well but this is most evident in the top panel with logarithmic scaling. While both the chosen equation and the numerically calculated \radF~trend to zero at high $k$, the behavior is slightly different. The emulator function declines quicker, at least for values of $n<1$. 

\section{Application to Profile Fitting}
\label{sec:fitting}

\subsection{Injection-Recovery Tests}
Our final equation is selected based on fitting to the radial Fourier Transform of the \ser\ profile; however, this does not necessarily translate into accuracy in the recovered properties of interest when fitting a galaxy (e.g., $n$, $r_e$). To ensure our chosen equation is accurate in real space, we perform a set of injection recovery tests using a ``perfect'' \ser~profile.

To begin, we implement this emulator for rendering in \texttt{pysersic}.\footnote{This is currently implemented in the 
branch \texttt{sersic\_fourier\_emulator} \href{https://github.com/pysersic/pysersic/tree/sersic\_fourier\_emulator}{here} and will be included in the next release of v0.2} Given the linearity of the Hankel transform, it is straightforward to scale the emulator \radFemul$~(k,n)$, which assumes a flux and radii of unity, to any arbitrary values during the fitting process. Our ``true'' profile will be rendered using a heavily over-sampled pixel space renderer. The entire image is over-sampled by a factor of 10 and the inner 8 pixels are over-sampled by an additional factor of 20. 
\begin{figure}
    \centering
    \includegraphics[width=0.99\columnwidth]{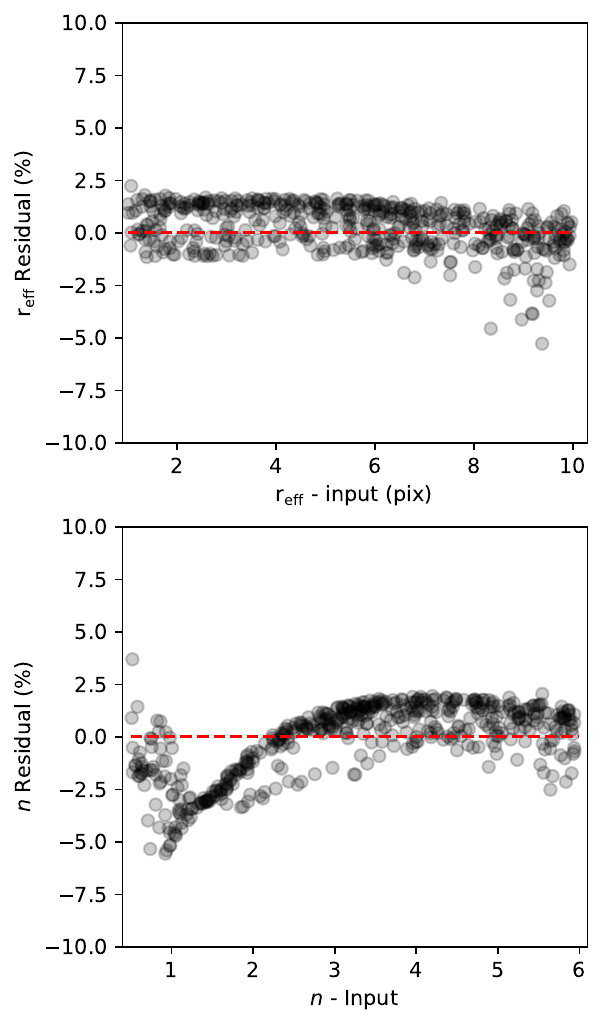}
    \caption{Residuals from the injection recovery test for $r_{\rm eff}$ and $n$ as a function of the input value for each parameter. This figure showcases the same data as shown in Fig.~\ref{fig:inj_recov}. The residuals for $r_{\rm eff}$ do not vary systematically however for $n$ we find the biggest disagreement at around $n \lesssim 1.5$ mirroring the results shown in Fig.~\ref{fig:eqn_resid}.}
    \label{fig:model_comp_resid}
\end{figure}

\begin{figure*}
    \centering
    \includegraphics[width=0.99\textwidth]{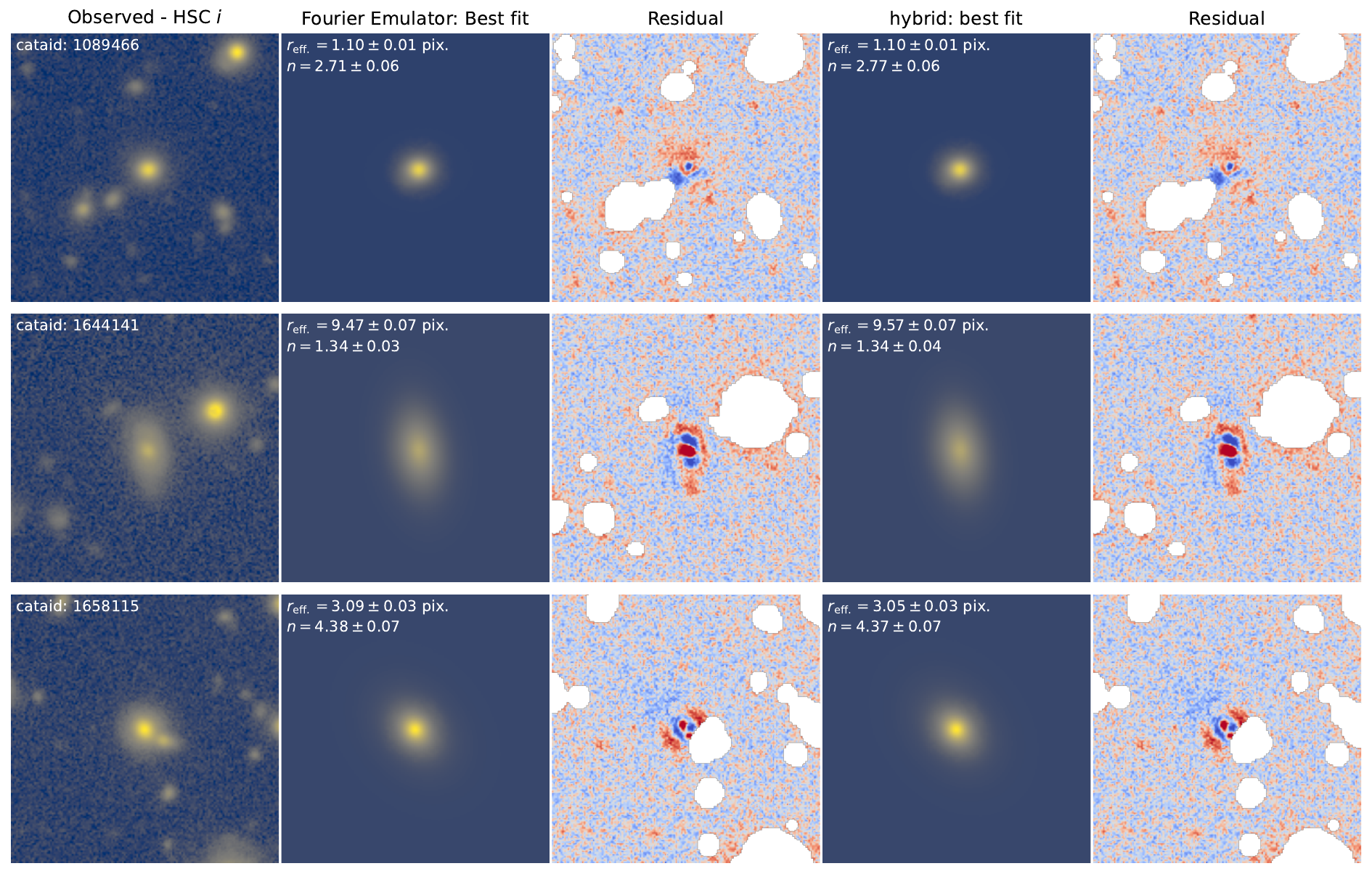}
    \caption{The results of profile fitting for three galaxies selected from the GAMA survey at $0.4<z<0.5$. We model the $i$ band images from HSC using \texttt{pysersic} and compare two different rendering methods: The default \texttt{hybrid} method based on using a mixture of Gaussians and the Fourier Emulator method presented in this work. For each method we show the best fit model, residual and the median and standard deviation of the posterior for the radius, $r_{\rm eff.}$, and index, $n$. All of the images are shown using an arcsinh stretch with the same scaling across all images of the same galaxy. Comparing the two rendering methods, it is difficult to distinguish the model images by eye and we find the recovered parameters to be very similar.}
    \label{fig:HSC_ind_gal}
\end{figure*}

We render a sample of 500 ``true'' \ser{} profiles onto a 200 pixel by 200 pixel image with $r_{\rm eff}$ randomly sampled between 1-10 pix, and $n$ between $0.5-6.0$. We additionally sample random values of the axis ratio uniformly between 0 and 0.9, position angle  uniformly between 0 and $\pi$ and central position in each coordinate following a Gaussian centered on the center of the image with a width of 0.5 pixels. The flux of each profile is assumed to be 100. We then find the best fit to each profile using the Fourier Emulator rendering scheme to ensure we achieve consistent results. To help stabilize the fitting process we add a small amount of Gaussian noise with an r.m.s of $10^{-4}$ per pixel. 

We note that in our initial optimization there was one source which was an outlier from the distribution, with best fit parameters $>20\%$ different from the input values. Upon further investigation we found that the best-fit position angle was exactly 0, the limit of our prior, when the input was $2.66$. This suggested an issue with the minimization. To confirm this we restarted the optimization using a different random seed and found the best-fit parameters were in much better alignment with the input values. We have chosen to only show this second minimization for this source and omit the initial run.

The results of our injection recovery tests are displayed in Figure~\ref{fig:inj_recov}. We find strong agreement between the input parameters and those recovered using the \ser~Fourier Emulator to render the profiles. For the central position, $x_c$ and $y_c$, ellipticity and position angle we find agreement to a level better than $10^{-3}$ The average difference between the input and recovered parameters is $<0.4\%$ and the scatter is $<1.7\%$ for $r_{\rm eff}$ and $n$ and $<0.5\%$ for the total flux. 

To further investigate the accuracy of the new rendering scheme we show the residuals of the recovery test for $n$ and $r_{\rm eff}$ as a function of the input values. The $r_{\rm eff}$ residuals do not show much structure, with the distribution around the true value remaining relatively consistent as a function of input radius. The residuals of $n$, do show systematic variation as a function of the input index. Most notably the emulator systematically underestimates the index by up to $5\%$ near $n=1$. For values of $n>3$ the index is instead overestimated by $1\%$. This mirrors the results shown in Fig.~\ref{fig:eqn_resid} which indicated the emulator was less accurate for profiles with $n\lesssim 1$.


\begin{figure*}
    \centering
    \includegraphics[width=0.99\textwidth]{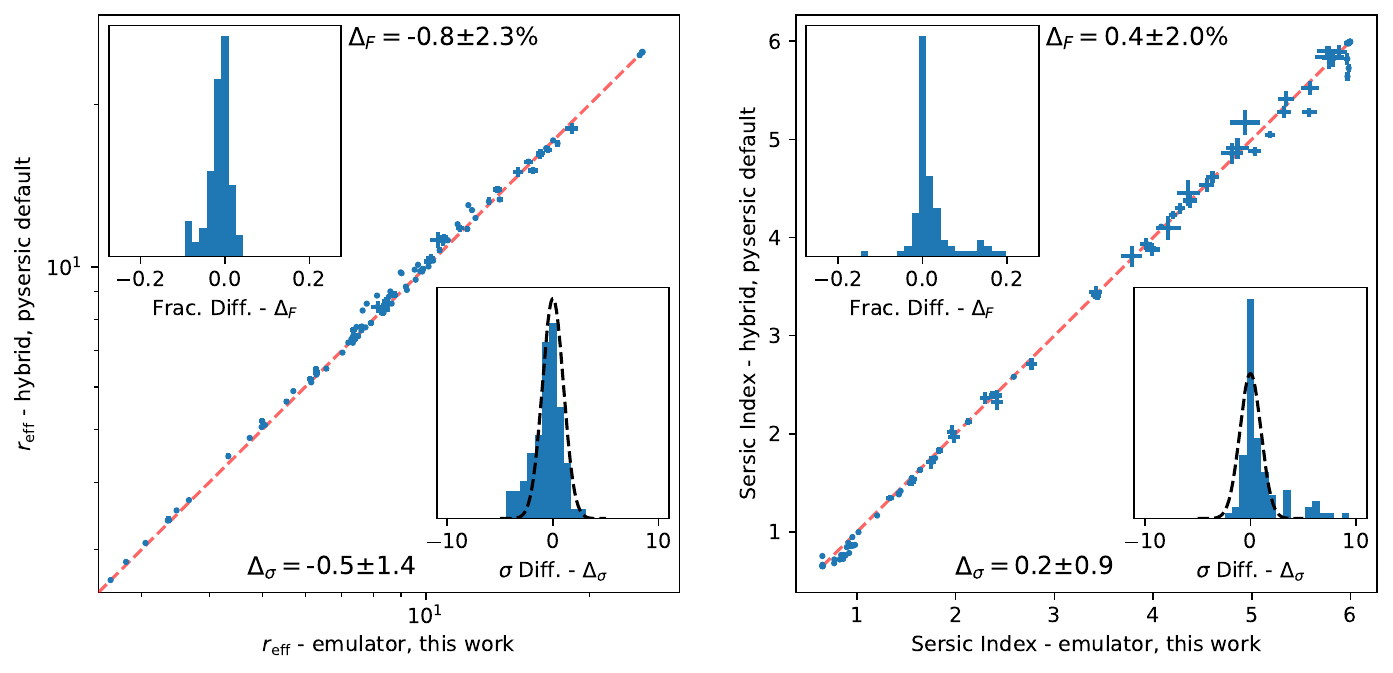}
    \caption{Comparing the recovered radius, $R_{\rm eff}$ and \ser~index,$n$, when using the \texttt{hybrid} and Fourier Emulator rendering methods for a sample of 100 galaxies with HSC imaging. The red line shows the one-to-one relation. Two insets are included for each parameter showcasing a histogram of the fractional difference in the upper left, and difference with respect to the uncertainties in the bottom right. For the latter the black dotted line showcases a standard normal distribution, which would be expected if there are no systematic differences between the two. We find the two rendering methods provide largely consistent results with the median fractional difference of $<1\%$ for both parameters with a scatter of $2.6\%$ for $r_{\rm eff.}$ and $4.9\%$ for $n$ along with the $\sigma$ differences largely following a standard normal distribution, with some outlier notably in $n$.}
    \label{fig:HSC_param}
\end{figure*}

\subsection{Real Galaxies}

Along with the injection-recovery tests we also wish to test how well this rendering technique performs in a realistic scenario. For this we compare to the \texttt{hybrid} method implemented in \texttt{pysersic}, which is the default. However both of these rendering methods are imperfect. Our goal with this test is not to assess absolute accuracy, as with the injection-recovery tests, but ensure we are able to provide consistent results with an established technique.

We select 100 galaxies from the Galaxy Mass And Assembly (GAMA) survey, Data Release 4, at $0.4<z<0.5$ in the G02 field ~\citep{Driver:2022}. This field overlaps with the HSC-Subaru Strategic Program imaging program which we will use to perform the morphological fits. Focusing on the $i$ band, science cutouts, weight maps and PSFs are downloaded from the public data release 2 using the \texttt{unagi} python package\footnote{https://github.com/dr-guangtou/unagi/tree/master}~\citep{Aihara:2019}. Our aim is to assess the accuracy and performance gain of the rendering method based on the \ser~Fourier emulator compared to the default \texttt{hybrid} method in \texttt{pysersic}.

For each galaxy we perform the morphological fits with \texttt{pysersic} twice: using the \texttt{hybrid} renderer and using the emulator function \radFemul~$(k,n)$. \texttt{hybrid} is the default rendering method in \texttt{pysersic} and implements the  method proposed in \citet{Lang:2020} using a series of Gaussians to represent the \ser~profile. In \texttt{pysersic} the amplitudes of each component are derived using the method described in ~\citet{Shajib:2019}. Following \citet{Lang:2020}, the majority of the Gaussian components are rendered in Fourier space while three with the largest width are rendered in observed space to address the issue of aliasing. For each renderer we use the same input data, prior and setup, using a student-T loss function, with $\nu=3$ and simultaneously fitting a flat sky background.

We perform inference using two methods. The first is stochastic variational inference (SVI)~\citep{Hoffman2013}. This method approximates the posterior distribution using a normalizing flow~\citep{DeCao:2020}. We also use a minimizer to estimate the maximum-a-posteriori (MAP) for each galaxy. It is generally faster but can be less accurate as it will only ever approximate the posterior. The next is MCMC sampling using the No U-Turn sampler implemented in \texttt{numpyro}~\citep{Hoffman:2014,Phan:2019} using 2 chains with $1,000$ warm-up and $1,000$ sampling steps. Before sampling we reparameterize the parameters using a multi-variate normal following \citet{Hoffman2019}. For these tests we use the default settings in \texttt{pysersic}. We ensure the resulting chains all have effective sample sizes $>300$ for all parameters and $\hat{r}<1.05$ for each renderer~\citep{Vehtari2021}. This removes 8 galaxies. Of the remaining galaxies the median effective sample size is 1148 and all values of $\hat{r}$ are $\leq1.01$. These tests are all performed on a 2023 Macbook Pro with an M2 Pro processor utilizing the CPU. 

The results of profile fitting for three example galaxies is shown in Figure~\ref{fig:HSC_ind_gal}. In this figure we can directly compare the results when using the Fourier Emulator and \texttt{hybrid} rendering methods. In general we find very little difference; the best fit models for both appear indistinguishable by eye. The recovered parameters are also well matched between the two methods. There are some differences but these are generally smaller than the uncertainties.

A quantitative comparison of the recovered parameters using the different methods is displayed in Figure~\ref{fig:HSC_param}. Similar to Fig.~\ref{fig:inj_recov} the recovered half-light radius and  \ser~index from the two different rendering methods are directly compared. There are two inset histograms: The upper left the distribution of fractional differences, denoted as $\Delta_f$ and the bottom right showcasing the difference with respect to the combined uncertainty, denoted $\Delta_\sigma$. The combined uncertainty is calculated by adding that measured from each method in quadrature, assuming they are independent.

We find good agreement between the two methods with the mean fractional difference, calculated using the biweight location~\citep{Beers1990}, to be $<1\%$ for both parameters. The scatter in $\Delta_F$, calculated using the biweight scale, is $2.6 \%$ for $r_{\rm eff.}$ and $4.9\%$ for $n$. We compare the distribution $\Delta_\sigma$ to a standard normal which one would expect to find if there were no systematic differences between the two. For $r_{\rm eff}$ we find that $\Delta_\sigma$ largely follows a normal distribution with a slight bias towards negative values. For $n$, the distribution of $\Delta_\sigma$ shows outliers at $>5\sigma$. These appear to be concentrated at small $n$, near 1, mirroring the results shown in Figure~\ref{fig:eqn_resid}. The distribution of $\Delta_{\sigma}$ for $n$ is centered at 0, with a width that appears to be slightly narrower than 1.

\begin{figure*}
    \centering
    \includegraphics[width=0.85\linewidth]{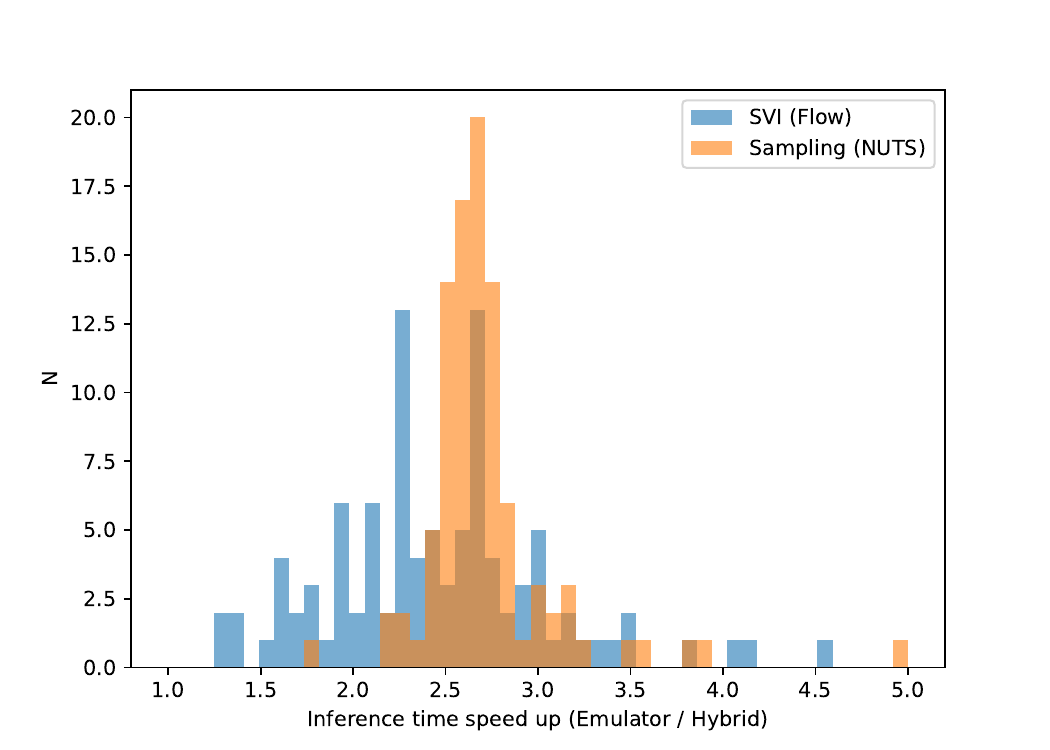}
    \caption{Comparing inference times in \texttt{pysersic} when using the default \texttt{hybrid}~\citep{Lang:2020} method compared to the \ser~Fourier Emulator methods described here. For both inference techniques the rendering scheme proposed in this work performs $2.5\times$ faster. This brings the median time to complete sampling, for example, from 64 seconds down to 23.5 seconds including all overheads. }
    \label{fig:infer_time_dist}
\end{figure*}

We investigate the difference in inference time between the \texttt{hybrid} and Fourier Emulator rendering methods in Figure~\ref{fig:infer_time_dist}. For both inference techniques, SVI and MCMC, we see a $2.5\times$ improvement in speed when using the Fourier Emulator rendering method. For MCMC sampling specifically this drops the median computational time from 64 sec down to 23.5 seconds. This timing includes all overheads including setup, compilation and well as saving the results. For the SVI method we find the median time decreases from 36.5 seconds to 15.3 seconds, when comparing the \texttt{hybrid} and Fourier Emulator rendering methods respectively. The ratio of median times of 2.4 is similar to that for MCMC sampling. For estimating the MAP, the median time ratio is 2.4, with a median time for 3.33 seconds when using the \texttt{hybrid} renderer and 1.45 seconds for the Fourier Emulator method described in this work.

\section{Summary and Discussion}
\label{sec:summ}

Galaxy profile fitting underpins much of the science performed with extragalactic surveys. The most common parameterization is the \ser~ profile, which can reasonably represent most galaxies but is difficult to work with analytically. Crucially, there is not a general known closed form Fourier transform. In this paper we have explored using symbolic regression to emulate the radial Fourier transform, or Hankel transform, of the \ser~profile. Our goal is to implement an accurate, computationally efficient, and differentiable rendering scheme for galaxy profile fitting.

To achieve this goal we numerically calculated \radF$(k,n)$ for a set of $k$, the spatial frequency coordinate and $n$ the \ser~ index, values using the \texttt{hankel} python package.~\citep{Murray:2019}. Analyzing the results of our symbolic regression fitting, performed using \texttt{pysr}~\citep{Cranmer2023}, in the context of profile fitting, we selected an equation that is both accurate and computationally efficient. The results of the symbolic regression fitting are shown in Figure~\ref{fig:loss_time} and the selected equation in Equation~\ref{eqn:emulator}. 

To test this new rendering method we perform two sets of tests. The first is a set of injection-recovery tests using a heavily over-sampled pixel rendering method as a ``perfect'' profile. We find the emulator provides consistent results with the average fractional difference in the inferred effective radius and \ser~index to be $<0.4\%$ for both parameters with scatter $<1.7\%$. However there is systematic variation in the accuracy of the index, with a $-5\%$ offset at $n\approx1$ and $2\%$ offset at $n>3$. The second is comparing this rendering method to the \texttt{hybrid} method implemented in \texttt{pysersic} applied to the same set of 100 galaxies in the HSC-SSP imaging. This is meant to test how it compares to established techniques in a real-world scenario. Using \texttt{pysersic}, we fit the same set of galaxies using the \ser~Fourier Emulator rendering method and the default \texttt{hybrid} method based on approximating the \ser~profile using a mixture of Gaussians. Similar to the injection-recovery tests we found consistent results when comparing the two rendering methods.

The systematic uncertainties imparted by the rendering methods proposed here are below the typical uncertainty of morphological catalogs. In order to assess the typical uncertainty we download two publicly available catalogs from \citet{VanderWel2012,George2024}. The \citet{VanderWel2012} catalog is based on HST data and shows a median fractional uncertainty of 8.9\% in the radius and 21.5\% in the index. \citet{George2024} uses data from the CLAUDS and HSC-SSP surveys and find a median fractional uncertainties of 7.5\% for radius and 11.0\% for the index. These are all much greater than the systematic uncertainty implied by our injection-recovery tests in Fig.~\ref{fig:inj_recov}. On a catalog level, the level of accuracy the emulator provides appears to be sufficient, however specific science goals may require more precision. 

Comparing the computational runtime, we find that the emulator rendering method performs the MCMC sampling a factor of 2.5 faster than the \texttt{hybrid} method with minimal loss in accuracy. This means on modern laptops MCMC sampling can be performed in less than 30 seconds per galaxy, which results in effective sample sizes $\gtrsim 1,000$. Alternative methods like variational inference can decrease the time required to $\sim 15$ seconds, and optimization to find a best-fitting point in $<1.5$ seconds. These improvements in computational efficiency are necessary to continue to scale scientific pipelines to upcoming surveys with ever increasing numbers of galaxies.

It is important to note that the equation we present in this work is not unique nor likely fully optimized for the task for profile fitting. The loss function we used is based on L2 distance in Fourier space and it is unclear how this metric translates to the accuracy of the rendered image in real space. While we have shown in Fig.~\ref{fig:inj_recov} that the equation presented in this work is able to accurately render \ser~profiles, it may be possible to define a different loss function which is directly related to the accuracy of the (final) rendered image. This would focus the optimization on a more practical benchmark. Additionally we have investigated the computational efficiency of the results equations post-hoc, but it may be possible to fold this information in during the fitting process as an additional value to optimize. Continued effort tweaking the symbolic regression procedure or using an entirely different approach will likely result in a more accurate and/or faster emulator.

Additionally, this approach is not without its limitations. First, it is an empirical approximation, and therefore is sensitive to inaccuracies and only valid for the range of parameters it has been trained on. The range of $n$ values we have used, $0.5<n<6$ encompasses the majority of real galaxies ~\citep[e.g.][]{Lange:2015,Mowla:2019} but it is common to see a higher $n$ limit, often up to $n=8$, used in profile fitting. Initially we attempted to find an emulator over the entire range of $0.5<n<8$ however we found the symbolic regression procedure had difficulty producing accurate results when using\radF~$(k,n)$ to perform profile fitting. There were systematic deviations on the scale of $\sim10\%$ for both high and low values of $n$, $n>6$ and $n<1$ respectively. To ensure an accurate equation which covers the vast majority of realistic galaxy profiles we therefore limited our training set, and thus the applicability of our emulator, to profiles with $n<6$. 

Another is aliasing, or ``wrapping around'' of extended sources near the edge of the image. This is a generic issue of rendering in Fourier space due to the periodic nature of the inverse FFT.~\citet{Lang:2020} introduces a method to overcome this issue using a series of Gaussians to represent the \ser~profile and render the largest components in real space. This avoids rendering most of the extended emission, which is the most susceptible to aliasing, in Fourier space therefore minimizing the effects of aliasing. Another option is to render on a larger grid and trim the edges back to the original shape. This simplistic approach simply removes the edges of the cutout where the aliasing is most prominent. However, doing so adds significant computational cost as larger grids need to be used.

In this initial implementation of the emulator rendering method we have not attempted any correction for this type of spatial aliasing. For \texttt{pysersic}, the focus is on modeling single galaxies that are almost always in the center of the cutout, so aliasing is often less of a concern. It may be possible to use the emulator to render in Fourier space up to a given spatial scale and perform the rest of the rendering in real space, similar to the \citet{Lang:2020} approach, but this will need to be implemented and tested.

In this work we have introduced a new method of rendering \ser~profiles by emulating its radial Fourier transform. We use symbolic regression to derive an empirical approximation shown in Eqn.~\ref{eqn:emulator} which is implemented in \texttt{pysersic} fitting code. We show that this new method performs measurements 2.5 times faster than the default rendering method with minimal loss in accuracy. This new type of rendering for galaxy profile fitting will help increase computational efficiency to allow pipelines to better scale to the ever increasing amount of data. Importantly, this method retains the differentiability necessary for modern optimization and Bayesian inference algorithms.

\section*{Acknowledgments}

TBM would like to thank the organizers of the JAXtronomy meeting at the Flatiron Institute where the idea for this paper originated. TBM was supported by a CIERA Postdoctoral Fellowship. 
 
This work used computing resources provided by Northwestern University and the Center for Interdisciplinary Exploration and Research in Astrophysics (CIERA). This research was supported in part through the computational resources and staff contributions provided for the Quest high performance computing facility at Northwestern University which is jointly supported by the Office of the Provost, the Office for Research, and Northwestern University Information Technology.

GAMA is a joint European-Australasian project based around a spectroscopic campaign using the Anglo-Australian Telescope. The GAMA input catalogue is based on data taken from the Sloan Digital Sky Survey and the UKIRT Infrared Deep Sky Survey. Complementary imaging of the GAMA regions is being obtained by a number of independent survey programmes including GALEX MIS, VST KiDS, VISTA VIKING, WISE, Herschel-ATLAS, GMRT and ASKAP providing UV to radio coverage. GAMA is funded by the STFC (UK), the ARC (Australia), the AAO, and the participating institutions. The GAMA website is https://www.gama-survey.org/ 

The Hyper Suprime-Cam Subaru Strategic Program (HSC-SSP) is led by the astronomical communities of Japan and Taiwan, and Princeton University.  The instrumentation and software were developed by the National Astronomical Observatory of Japan (NAOJ), the Kavli Institute for the Physics and Mathematics of the Universe (Kavli IPMU), the University of Tokyo, the High Energy Accelerator Research Organization (KEK), the Academia Sinica Institute for Astronomy and Astrophysics in Taiwan (ASIAA), and Princeton University.  The survey was made possible by funding contributed by the Ministry of Education, Culture, Sports, Science and Technology (MEXT), the Japan Society for the Promotion of Science (JSPS),  (Japan Science and Technology Agency (JST),  the Toray Science Foundation, NAOJ, Kavli IPMU, KEK, ASIAA,  and Princeton University.

\bibliographystyle{apj}

\bibliography{paper}

@ARTICLE{Vehtari2021,
       author = {{Vehtari}, Aki and {Gelman}, Andrew and {Simpson}, Daniel and {Carpenter}, Bob and {B{\"u}rkner}, Paul-Christian},
        title = "{Rank-normalization, folding, and localization: An improved R-hat for assessing convergence of MCMC (with Discussion)}",
      journal = {Bayesian Analysis},
         year = 2021,
        month = jun,
       volume = {16},
       number = {2},
        pages = {667-718},
          doi = {10.1214/20-BA1221},
archivePrefix = {arXiv},
       eprint = {1903.08008},
 primaryClass = {stat.CO},
       adsurl = {https://ui.adsabs.harvard.edu/abs/2021BayAn..16..667V}
}

@ARTICLE{VanderWel2012,
       author = {{van der Wel}, A. and {Bell}, E.~F. and {H{\"a}ussler}, B. and {McGrath}, E.~J. and {Chang}, Yu-Yen and {Guo}, Yicheng and {McIntosh}, D.~H. and {Rix}, H.-W. and {Barden}, M. and {Cheung}, E. and {Faber}, S.~M. and {Ferguson}, H.~C. and {Galametz}, A. and {Grogin}, N.~A. and {Hartley}, W. and {Kartaltepe}, J.~S. and {Kocevski}, D.~D. and {Koekemoer}, A.~M. and {Lotz}, J. and {Mozena}, M. and {Peth}, M.~A. and {Peng}, Chien Y.},
        title = "{Structural Parameters of Galaxies in CANDELS}",
      journal = {\apjs},
         year = 2012,
        month = dec,
       volume = {203},
       number = {2},
          eid = {24},
        pages = {24},
          doi = {10.1088/0067-0049/203/2/24},
archivePrefix = {arXiv},
       eprint = {1211.6954},
 primaryClass = {astro-ph.CO},
       adsurl = {https://ui.adsabs.harvard.edu/abs/2012ApJS..203...24V}
}

@ARTICLE{George2024,
       author = {{George}, Angelo and {Damjanov}, Ivana and {Sawicki}, Marcin and {Arnouts}, St{\'e}phane and {Desprez}, Guillaume and {Gwyn}, Stephen and {Picouet}, Vincent and {Birrer}, Simon and {Silverman}, John},
        title = "{Two rest-frame wavelength measurements of galaxy sizes at z < 1: the evolutionary effects of emerging bulges and quenched newcomers}",
      journal = {\mnras},
         year = 2024,
        month = mar,
       volume = {528},
       number = {3},
        pages = {4797-4828},
          doi = {10.1093/mnras/stae154},
archivePrefix = {arXiv},
       eprint = {2401.06842},
 primaryClass = {astro-ph.GA},
       adsurl = {https://ui.adsabs.harvard.edu/abs/2024MNRAS.528.4797G}
}

@ARTICLE{Bernardi2013,
       author = {{Bernardi}, M. and {Meert}, A. and {Sheth}, R.~K. and {Vikram}, V. and {Huertas-Company}, M. and {Mei}, S. and {Shankar}, F.},
        title = "{The massive end of the luminosity and stellar mass functions: dependence on the fit to the light profile}",
      journal = {\mnras},
         year = 2013,
        month = nov,
       volume = {436},
       number = {1},
        pages = {697-704},
          doi = {10.1093/mnras/stt1607},
archivePrefix = {arXiv},
       eprint = {1304.7778},
 primaryClass = {astro-ph.CO},
       adsurl = {https://ui.adsabs.harvard.edu/abs/2013MNRAS.436..697B}
}

@software{Baldwin2024,
       author = {{Baldwin}, James O. and {Nelson}, Erica and {Johnson}, Benjamin D. and {Oesch}, Pascal A. and {Tacchella}, Sandro and {Illingworth}, Garth D. and {Gibson}, Justus and {Hartley}, Abby},
        title = "{forcepho: Generative modeling galaxy photometry for JWST}",
 howpublished = {Astrophysics Source Code Library, record ascl:2410.006},
         year = 2024,
        month = oct,
          eid = {ascl:2410.006},
       adsurl = {https://ui.adsabs.harvard.edu/abs/2024ascl.soft10006B}
}

@article{Hoffman2013,
  title={Stochastic variational inference},
  author={Hoffman, Matthew D and Blei, David M and Wang, Chong and Paisley, John},
  journal={the Journal of machine Learning research},
  volume={14},
  number={1},
  pages={1303--1347},
  year={2013},
  publisher={JMLR. org}
}

@ARTICLE{Beers1990,
       author = {{Beers}, Timothy C. and {Flynn}, Kevin and {Gebhardt}, Karl},
        title = "{Measures of Location and Scale for Velocities in Clusters of Galaxies---A Robust Approach}",
      journal = {\aj},
         year = 1990,
        month = jul,
       volume = {100},
        pages = {32},
          doi = {10.1086/115487},
       adsurl = {https://ui.adsabs.harvard.edu/abs/1990AJ....100...32B}
}

@ARTICLE{Hoffman2019,
       author = {{Hoffman}, Matthew and {Sountsov}, Pavel and {Dillon}, Joshua V. and {Langmore}, Ian and {Tran}, Dustin and {Vasudevan}, Srinivas},
        title = "{NeuTra-lizing Bad Geometry in Hamiltonian Monte Carlo Using Neural Transport}",
      journal = {arXiv e-prints},
         year = 2019,
        month = mar,
          eid = {arXiv:1903.03704},
        pages = {arXiv:1903.03704},
          doi = {10.48550/arXiv.1903.03704},
archivePrefix = {arXiv},
       eprint = {1903.03704},
 primaryClass = {stat.CO},
       adsurl = {https://ui.adsabs.harvard.edu/abs/2019arXiv190303704H}
}

@ARTICLE{Aihara:2019,
       author = {{Aihara}, Hiroaki and {AlSayyad}, Yusra and {Ando}, Makoto and {Armstrong}, Robert and {Bosch}, James and {Egami}, Eiichi and {Furusawa}, Hisanori and {Furusawa}, Junko and {Goulding}, Andy and {Harikane}, Yuichi and {Hikage}, Chiaki and {Ho}, Paul T.~P. and {Hsieh}, Bau-Ching and {Huang}, Song and {Ikeda}, Hiroyuki and {Imanishi}, Masatoshi and {Ito}, Kei and {Iwata}, Ikuru and {Jaelani}, Anton T. and {Kakuma}, Ryota and {Kawana}, Kojiro and {Kikuta}, Satoshi and {Kobayashi}, Umi and {Koike}, Michitaro and {Komiyama}, Yutaka and {Li}, Xiangchong and {Liang}, Yongming and {Lin}, Yen-Ting and {Luo}, Wentao and {Lupton}, Robert and {Lust}, Nate B. and {MacArthur}, Lauren A. and {Matsuoka}, Yoshiki and {Mineo}, Sogo and {Miyatake}, Hironao and {Miyazaki}, Satoshi and {More}, Surhud and {Murata}, Ryoma and {Namiki}, Shigeru V. and {Nishizawa}, Atsushi J. and {Oguri}, Masamune and {Okabe}, Nobuhiro and {Okamoto}, Sakurako and {Okura}, Yuki and {Ono}, Yoshiaki and {Onodera}, Masato and {Onoue}, Masafusa and {Osato}, Ken and {Ouchi}, Masami and {Shibuya}, Takatoshi and {Strauss}, Michael A. and {Sugiyama}, Naoshi and {Suto}, Yasushi and {Takada}, Masahiro and {Takagi}, Yuhei and {Takata}, Tadafumi and {Takita}, Satoshi and {Tanaka}, Masayuki and {Terai}, Tsuyoshi and {Toba}, Yoshiki and {Uchiyama}, Hisakazu and {Utsumi}, Yousuke and {Wang}, Shiang-Yu and {Wang}, Wenting and {Yamada}, Yoshihiko},
        title = "{Second data release of the Hyper Suprime-Cam Subaru Strategic Program}",
      journal = {\pasj},
         year = 2019,
        month = dec,
       volume = {71},
       number = {6},
          eid = {114},
        pages = {114},
          doi = {10.1093/pasj/psz103},
archivePrefix = {arXiv},
       eprint = {1905.12221},
 primaryClass = {astro-ph.IM},
       adsurl = {https://ui.adsabs.harvard.edu/abs/2019PASJ...71..114A}
}

@ARTICLE{Driver:2022,
       author = {{Driver}, Simon P. and {Bellstedt}, Sabine and {Robotham}, Aaron S.~G. and {Baldry}, Ivan K. and {Davies}, Luke J. and {Liske}, Jochen and {Obreschkow}, Danail and {Taylor}, Edward N. and {Wright}, Angus H. and {Alpaslan}, Mehmet and {Bamford}, Steven P. and {Bauer}, Amanda E. and {Bland-Hawthorn}, Joss and {Bilicki}, Maciej and {Bravo}, Mat{\'\i}as and {Brough}, Sarah and {Casura}, Sarah and {Cluver}, Michelle E. and {Colless}, Matthew and {Conselice}, Christopher J. and {Croom}, Scott M. and {de Jong}, Jelte and {D'Eugenio}, Franceso and {De Propris}, Roberto and {Dogruel}, Burak and {Drinkwater}, Michael J. and {Dvornik}, Andrej and {Farrow}, Daniel J. and {Frenk}, Carlos S. and {Giblin}, Benjamin and {Graham}, Alister W. and {Grootes}, Meiert W. and {Gunawardhana}, Madusha L.~P. and {Hashemizadeh}, Abdolhosein and {H{\"a}u{\ss}ler}, Boris and {Heymans}, Catherine and {Hildebrandt}, Hendrik and {Holwerda}, Benne W. and {Hopkins}, Andrew M. and {Jarrett}, Tom H. and {Heath Jones}, D. and {Kelvin}, Lee S. and {Koushan}, Soheil and {Kuijken}, Konrad and {Lara-L{\'o}pez}, Maritza A. and {Lange}, Rebecca and {L{\'o}pez-S{\'a}nchez}, {\'A}ngel R. and {Loveday}, Jon and {Mahajan}, Smriti and {Meyer}, Martin and {Moffett}, Amanda J. and {Napolitano}, Nicola R. and {Norberg}, Peder and {Owers}, Matt S. and {Radovich}, Mario and {Raouf}, Mojtaba and {Peacock}, John A. and {Phillipps}, Steven and {Pimbblet}, Kevin A. and {Popescu}, Cristina and {Said}, Khaled and {Sansom}, Anne E. and {Seibert}, Mark and {Sutherland}, Will J. and {Thorne}, Jessica E. and {Tuffs}, Richard J. and {Turner}, Ryan and {van der Wel}, Arjen and {van Kampen}, Eelco and {Wilkins}, Steve M.},
        title = "{Galaxy And Mass Assembly (GAMA): Data Release 4 and the z < 0.1 total and z < 0.08 morphological galaxy stellar mass functions}",
      journal = {\mnras},
         year = 2022,
        month = jun,
       volume = {513},
       number = {1},
        pages = {439-467},
          doi = {10.1093/mnras/stac472},
archivePrefix = {arXiv},
       eprint = {2203.08539},
 primaryClass = {astro-ph.GA},
       adsurl = {https://ui.adsabs.harvard.edu/abs/2022MNRAS.513..439D}
}

@article{Angelis2023,
  title={Artificial intelligence in physical sciences: Symbolic regression trends and perspectives},
  author={Angelis, Dimitrios and Sofos, Filippos and Karakasidis, Theodoros E},
  journal={Archives of Computational Methods in Engineering},
  volume={30},
  number={6},
  pages={3845--3865},
  year={2023},
  publisher={Springer}
}

@ARTICLE{chen2024,
       author = {{Chen}, Mi and {S. de Souza}, Rafael and {Xu}, Quanfeng and {Shen}, Shiyin and {Chies-Santos}, Ana L. and {Ye}, Renhao and {Canossa-Gosteinski}, Marco A. and {Cong}, Yanping},
        title = "{Galmoss: A package for GPU-accelerated galaxy profile fitting}",
      journal = {Astronomy and Computing},
         year = 2024,
        month = apr,
       volume = {47},
          eid = {100825},
        pages = {100825},
          doi = {10.1016/j.ascom.2024.100825},
archivePrefix = {arXiv},
       eprint = {2404.07780},
 primaryClass = {astro-ph.GA},
       adsurl = {https://ui.adsabs.harvard.edu/abs/2024A&C....4700825C}
}

@article{Murray:2019, doi = {10.21105/joss.01397}, url = {https://doi.org/10.21105/joss.01397}, year = {2019}, publisher = {The Open Journal}, volume = {4}, number = {37}, pages = {1397}, author = {Steven G. Murray and Francis J. Poulin}, title = {hankel: A Python library for performing simple and accurate Hankel transformations}, journal = {Journal of Open Source Software} }

@article{Ogata:2005,
  title={A numerical integration formula based on the Bessel functions},
  author={Ogata, Hidenori},
  journal={Publications of the Research Institute for Mathematical Sciences},
  volume={41},
  number={4},
  pages={949--970},
  year={2005}
}

@ARTICLE{Cranmer2023,
       author = {{Cranmer}, Miles},
        title = "{Interpretable Machine Learning for Science with PySR and SymbolicRegression.jl}",
      journal = {arXiv e-prints},
         year = 2023,
        month = may,
          eid = {arXiv:2305.01582},
        pages = {arXiv:2305.01582},
          doi = {10.48550/arXiv.2305.01582},
archivePrefix = {arXiv},
       eprint = {2305.01582},
 primaryClass = {astro-ph.IM},
       adsurl = {https://ui.adsabs.harvard.edu/abs/2023arXiv230501582C}
}

@ARTICLE{Li:2023,
       author = {{Li}, Jiaxuan and {Greene}, Jenny E. and {Greco}, Johnny P. and {Huang}, Song and {Melchior}, Peter and {Beaton}, Rachael and {Casey}, Kirsten and {Danieli}, Shany and {Goulding}, Andy and {Joseph}, Remy and {Kado-Fong}, Erin and {Kim}, Ji Hoon and {MacArthur}, Lauren A.},
        title = "{Beyond Ultra-diffuse Galaxies. I. Mass-Size Outliers among the Satellites of Milky Way Analogs}",
      journal = {\apj},
         year = 2023,
        month = sep,
       volume = {955},
       number = {1},
          eid = {1},
        pages = {1},
          doi = {10.3847/1538-4357/ace829},
archivePrefix = {arXiv},
       eprint = {2210.14994},
 primaryClass = {astro-ph.GA},
       adsurl = {https://ui.adsabs.harvard.edu/abs/2023ApJ...955....1L}
}

@ARTICLE{Tan2024,
       author = {{Tan}, Qing-Hua and {Daddi}, Emanuele and {de Souza Magalh{\~a}es}, Victor and {G{\'o}mez-Guijarro}, Carlos and {Pety}, J{\'e}r{\^o}me and {Kalita}, Boris S. and {Elbaz}, David and {Liu}, Zhaoxuan and {Magnelli}, Benjamin and {Puglisi}, Annagrazia and {Rujopakarn}, Wiphu and {Silverman}, John D. and {Valentino}, Francesco and {Zhang}, Shao-Bo},
        title = "{Fitting pseudo-S{\'e}rsic (Spergel) light profiles to galaxies in interferometric data: The excellence of the u{\ensuremath{\upsilon}}-plane}",
      journal = {\aap},
         year = 2024,
        month = apr,
       volume = {684},
          eid = {A23},
        pages = {A23},
          doi = {10.1051/0004-6361/202347255},
archivePrefix = {arXiv},
       eprint = {2312.05425},
 primaryClass = {astro-ph.GA},
       adsurl = {https://ui.adsabs.harvard.edu/abs/2024A&A...684A..23T}
}

@ARTICLE{Spergel2010,
       author = {{Spergel}, David N.},
        title = "{Analytical Galaxy Profiles for Photometric and Lensing Analysis}",
      journal = {\apjs},
         year = 2010,
        month = nov,
       volume = {191},
       number = {1},
        pages = {58-65},
          doi = {10.1088/0067-0049/191/1/58},
archivePrefix = {arXiv},
       eprint = {1007.3248},
 primaryClass = {astro-ph.CO},
       adsurl = {https://ui.adsabs.harvard.edu/abs/2010ApJS..191...58S}
}

@software{Lang:2016,
       author = {{Lang}, Dustin and {Hogg}, David W. and {Mykytyn}, David},
        title = "{The Tractor: Probabilistic astronomical source detection and measurement}",
 howpublished = {Astrophysics Source Code Library, record ascl:1604.008},
         year = 2016,
        month = apr,
          eid = {ascl:1604.008},
       adsurl = {https://ui.adsabs.harvard.edu/abs/2016ascl.soft04008L}
}

@ARTICLE{Hogg2013,
       author = {{Hogg}, David W. and {Lang}, Dustin},
        title = "{Replacing Standard Galaxy Profiles with Mixtures of Gaussians}",
      journal = {\pasp},
         year = 2013,
        month = jun,
       volume = {125},
       number = {928},
        pages = {719},
          doi = {10.1086/671228},
archivePrefix = {arXiv},
       eprint = {1210.6563},
 primaryClass = {astro-ph.IM},
       adsurl = {https://ui.adsabs.harvard.edu/abs/2013PASP..125..719H}
}

@ARTICLE{Rowe:2015,
       author = {{Rowe}, B.~T.~P. and {Jarvis}, M. and {Mandelbaum}, R. and {Bernstein}, G.~M. and {Bosch}, J. and {Simet}, M. and {Meyers}, J.~E. and {Kacprzak}, T. and {Nakajima}, R. and {Zuntz}, J. and {Miyatake}, H. and {Dietrich}, J.~P. and {Armstrong}, R. and {Melchior}, P. and {Gill}, M.~S.~S.},
        title = "{GALSIM: The modular galaxy image simulation toolkit}",
      journal = {Astronomy and Computing},
         year = 2015,
        month = apr,
       volume = {10},
        pages = {121-150},
          doi = {10.1016/j.ascom.2015.02.002},
archivePrefix = {arXiv},
       eprint = {1407.7676},
 primaryClass = {astro-ph.IM},
       adsurl = {https://ui.adsabs.harvard.edu/abs/2015A&C....10..121R}
}

@ARTICLE{Stone:2023,
       author = {{Stone}, Connor J. and {Courteau}, St{\'e}phane and {Cuillandre}, Jean-Charles and {Hezaveh}, Yashar and {Perreault-Levasseur}, Laurence and {Arora}, Nikhil},
        title = "{ASTROPHOT: fitting everything everywhere all at once in astronomical images}",
      journal = {\mnras},
         year = 2023,
        month = nov,
       volume = {525},
       number = {4},
        pages = {6377-6393},
          doi = {10.1093/mnras/stad2477},
archivePrefix = {arXiv},
       eprint = {2308.01957},
 primaryClass = {astro-ph.IM},
       adsurl = {https://ui.adsabs.harvard.edu/abs/2023MNRAS.525.6377S}
}

@ARTICLE{Pasha:2023,
       author = {{Pasha}, Imad and {Miller}, Tim B.},
        title = "{pysersic: A Python package for determining galaxy structural properties via Bayesian inference, accelerated with jax}",
      journal = {The Journal of Open Source Software},
         year = 2023,
        month = sep,
       volume = {8},
       number = {89},
          eid = {5703},
        pages = {5703},
          doi = {10.21105/joss.05703},
archivePrefix = {arXiv},
       eprint = {2306.05454},
 primaryClass = {astro-ph.GA},
       adsurl = {https://ui.adsabs.harvard.edu/abs/2023JOSS....8.5703P}
}

@ARTICLE{Allen:2024,
       author = {{Allen}, Natalie and {Oesch}, Pascal A. and {Toft}, Sune and {Matharu}, Jasleen and {McPartland}, Conor J.~R. and {Weibel}, Andrea and {Brammer}, Gabe and {Bowler}, Rebecca A.~A. and {Ito}, Kei and {Gottumukkala}, Rashmi and {Rizzo}, Francesca and {Valentino}, Francesco and {Varadaraj}, Rohan G. and {Weaver}, John R. and {Whitaker}, Katherine E.},
        title = "{Galaxy Size and Mass Build-up in the First 2 Gyrs of Cosmic History from Multi-Wavelength JWST NIRCam Imaging}",
      journal = {arXiv e-prints},
         year = 2024,
        month = oct,
          eid = {arXiv:2410.16354},
        pages = {arXiv:2410.16354},
          doi = {10.48550/arXiv.2410.16354},
archivePrefix = {arXiv},
       eprint = {2410.16354},
 primaryClass = {astro-ph.GA},
       adsurl = {https://ui.adsabs.harvard.edu/abs/2024arXiv241016354A}
}

@ARTICLE{Ward:2024,
       author = {{Ward}, Ethan and {de la Vega}, Alexander and {Mobasher}, Bahram and {McGrath}, Elizabeth J. and {Iyer}, Kartheik G. and {Calabr{\`o}}, Antonello and {Costantin}, Luca and {Dickinson}, Mark and {Holwerda}, Benne W. and {Huertas-Company}, Marc and {Hirschmann}, Michaela and {Lucas}, Ray A. and {Pandya}, Viraj and {Wilkins}, Stephen M. and {Yung}, L.~Y. Aaron and {Arrabal Haro}, Pablo and {Bagley}, Micaela B. and {Finkelstein}, Steven L. and {Kartaltepe}, Jeyhan S. and {Koekemoer}, Anton M. and {Papovich}, Casey and {Pirzkal}, Nor},
        title = "{Evolution of the Size{\textendash}Mass Relation of Star-forming Galaxies Since z = 5.5 Revealed by CEERS}",
      journal = {\apj},
         year = 2024,
        month = feb,
       volume = {962},
       number = {2},
          eid = {176},
        pages = {176},
          doi = {10.3847/1538-4357/ad20ed},
archivePrefix = {arXiv},
       eprint = {2311.02162},
 primaryClass = {astro-ph.GA},
       adsurl = {https://ui.adsabs.harvard.edu/abs/2024ApJ...962..176W}
}

@ARTICLE{Dey2019,
       author = {{Dey}, Arjun and {Schlegel}, David J. and {Lang}, Dustin and {Blum}, Robert and {Burleigh}, Kaylan and {Fan}, Xiaohui and {Findlay}, Joseph R. and {Finkbeiner}, Doug and {Herrera}, David and {Juneau}, St{\'e}phanie and {Landriau}, Martin and {Levi}, Michael and {McGreer}, Ian and {Meisner}, Aaron and {Myers}, Adam D. and {Moustakas}, John and {Nugent}, Peter and {Patej}, Anna and {Schlafly}, Edward F. and {Walker}, Alistair R. and {Valdes}, Francisco and {Weaver}, Benjamin A. and {Y{\`e}che}, Christophe and {Zou}, Hu and {Zhou}, Xu and {Abareshi}, Behzad and {Abbott}, T.~M.~C. and {Abolfathi}, Bela and {Aguilera}, C. and {Alam}, Shadab and {Allen}, Lori and {Alvarez}, A. and {Annis}, James and {Ansarinejad}, Behzad and {Aubert}, Marie and {Beechert}, Jacqueline and {Bell}, Eric F. and {BenZvi}, Segev Y. and {Beutler}, Florian and {Bielby}, Richard M. and {Bolton}, Adam S. and {Brice{\~n}o}, C{\'e}sar and {Buckley-Geer}, Elizabeth J. and {Butler}, Karen and {Calamida}, Annalisa and {Carlberg}, Raymond G. and {Carter}, Paul and {Casas}, Ricard and {Castander}, Francisco J. and {Choi}, Yumi and {Comparat}, Johan and {Cukanovaite}, Elena and {Delubac}, Timoth{\'e}e and {DeVries}, Kaitlin and {Dey}, Sharmila and {Dhungana}, Govinda and {Dickinson}, Mark and {Ding}, Zhejie and {Donaldson}, John B. and {Duan}, Yutong and {Duckworth}, Christopher J. and {Eftekharzadeh}, Sarah and {Eisenstein}, Daniel J. and {Etourneau}, Thomas and {Fagrelius}, Parker A. and {Farihi}, Jay and {Fitzpatrick}, Mike and {Font-Ribera}, Andreu and {Fulmer}, Leah and {G{\"a}nsicke}, Boris T. and {Gaztanaga}, Enrique and {George}, Koshy and {Gerdes}, David W. and {Gontcho}, Satya Gontcho A. and {Gorgoni}, Claudio and {Green}, Gregory and {Guy}, Julien and {Harmer}, Diane and {Hernandez}, M. and {Honscheid}, Klaus and {Huang}, Lijuan Wendy and {James}, David J. and {Jannuzi}, Buell T. and {Jiang}, Linhua and {Joyce}, Richard and {Karcher}, Armin and {Karkar}, Sonia and {Kehoe}, Robert and {Kneib}, Jean-Paul and {Kueter-Young}, Andrea and {Lan}, Ting-Wen and {Lauer}, Tod R. and {Le Guillou}, Laurent and {Le Van Suu}, Auguste and {Lee}, Jae Hyeon and {Lesser}, Michael and {Perreault Levasseur}, Laurence and {Li}, Ting S. and {Mann}, Justin L. and {Marshall}, Robert and {Mart{\'\i}nez-V{\'a}zquez}, C.~E. and {Martini}, Paul and {du Mas des Bourboux}, H{\'e}lion and {McManus}, Sean and {Meier}, Tobias Gabriel and {M{\'e}nard}, Brice and {Metcalfe}, Nigel and {Mu{\~n}oz-Guti{\'e}rrez}, Andrea and {Najita}, Joan and {Napier}, Kevin and {Narayan}, Gautham and {Newman}, Jeffrey A. and {Nie}, Jundan and {Nord}, Brian and {Norman}, Dara J. and {Olsen}, Knut A.~G. and {Paat}, Anthony and {Palanque-Delabrouille}, Nathalie and {Peng}, Xiyan and {Poppett}, Claire L. and {Poremba}, Megan R. and {Prakash}, Abhishek and {Rabinowitz}, David and {Raichoor}, Anand and {Rezaie}, Mehdi and {Robertson}, A.~N. and {Roe}, Natalie A. and {Ross}, Ashley J. and {Ross}, Nicholas P. and {Rudnick}, Gregory and {Safonova}, Sasha and {Saha}, Abhijit and {S{\'a}nchez}, F. Javier and {Savary}, Elodie and {Schweiker}, Heidi and {Scott}, Adam and {Seo}, Hee-Jong and {Shan}, Huanyuan and {Silva}, David R. and {Slepian}, Zachary and {Soto}, Christian and {Sprayberry}, David and {Staten}, Ryan and {Stillman}, Coley M. and {Stupak}, Robert J. and {Summers}, David L. and {Sien Tie}, Suk and {Tirado}, H. and {Vargas-Maga{\~n}a}, Mariana and {Vivas}, A. Katherina and {Wechsler}, Risa H. and {Williams}, Doug and {Yang}, Jinyi and {Yang}, Qian and {Yapici}, Tolga and {Zaritsky}, Dennis and {Zenteno}, A. and {Zhang}, Kai and {Zhang}, Tianmeng and {Zhou}, Rongpu and {Zhou}, Zhimin},
        title = "{Overview of the DESI Legacy Imaging Surveys}",
      journal = {\aj},
         year = 2019,
        month = may,
       volume = {157},
       number = {5},
          eid = {168},
        pages = {168},
          doi = {10.3847/1538-3881/ab089d},
archivePrefix = {arXiv},
       eprint = {1804.08657},
 primaryClass = {astro-ph.IM},
       adsurl = {https://ui.adsabs.harvard.edu/abs/2019AJ....157..168D}
}

@ARTICLE{Weaver2022,
       author = {{Weaver}, J.~R. and {Kauffmann}, O.~B. and {Ilbert}, O. and {McCracken}, H.~J. and {Moneti}, A. and {Toft}, S. and {Brammer}, G. and {Shuntov}, M. and {Davidzon}, I. and {Hsieh}, B.~C. and {Laigle}, C. and {Anastasiou}, A. and {Jespersen}, C.~K. and {Vinther}, J. and {Capak}, P. and {Casey}, C.~M. and {McPartland}, C.~J.~R. and {Milvang-Jensen}, B. and {Mobasher}, B. and {Sanders}, D.~B. and {Zalesky}, L. and {Arnouts}, S. and {Aussel}, H. and {Dunlop}, J.~S. and {Faisst}, A. and {Franx}, M. and {Furtak}, L.~J. and {Fynbo}, J.~P.~U. and {Gould}, K.~M.~L. and {Greve}, T.~R. and {Gwyn}, S. and {Kartaltepe}, J.~S. and {Kashino}, D. and {Koekemoer}, A.~M. and {Kokorev}, V. and {Le F{\`e}vre}, O. and {Lilly}, S. and {Masters}, D. and {Magdis}, G. and {Mehta}, V. and {Peng}, Y. and {Riechers}, D.~A. and {Salvato}, M. and {Sawicki}, M. and {Scarlata}, C. and {Scoville}, N. and {Shirley}, R. and {Silverman}, J.~D. and {Sneppen}, A. and {Smolc̆i{\'c}}, V. and {Steinhardt}, C. and {Stern}, D. and {Tanaka}, M. and {Taniguchi}, Y. and {Teplitz}, H.~I. and {Vaccari}, M. and {Wang}, W. -H. and {Zamorani}, G.},
        title = "{COSMOS2020: A Panchromatic View of the Universe to z{\ensuremath{\sim}}10 from Two Complementary Catalogs}",
      journal = {\apjs},
         year = 2022,
        month = jan,
       volume = {258},
       number = {1},
          eid = {11},
        pages = {11},
          doi = {10.3847/1538-4365/ac3078},
archivePrefix = {arXiv},
       eprint = {2110.13923},
 primaryClass = {astro-ph.GA},
       adsurl = {https://ui.adsabs.harvard.edu/abs/2022ApJS..258...11W}
}

@ARTICLE{devaucouleurs1948,
       author = {{de Vaucouleurs}, Gerard},
        title = "{Recherches sur les Nebuleuses Extragalactiques}",
      journal = {Annales d'Astrophysique},
         year = 1948,
        month = jan,
       volume = {11},
        pages = {247},
       adsurl = {https://ui.adsabs.harvard.edu/abs/1948AnAp...11..247D}
}

@ARTICLE{Sersic1963,
       author = {{S{\'e}rsic}, J.~L.},
        title = "{Influence of the atmospheric and instrumental dispersion on the brightness distribution in a galaxy}",
      journal = {Boletin de la Asociacion Argentina de Astronomia La Plata Argentina},
         year = 1963,
        month = feb,
       volume = {6},
        pages = {41-43},
       adsurl = {https://ui.adsabs.harvard.edu/abs/1963BAAA....6...41S}
}

@ARTICLE{Lang:2020,
       author = {{Lang}, Dustin},
        title = "{A hybrid Fourier--Real Gaussian Mixture method for fast galaxy--PSF convolution}",
      journal = {arXiv e-prints},
         year = 2020,
        month = dec,
          eid = {arXiv:2012.15797},
        pages = {arXiv:2012.15797},
          doi = {10.48550/arXiv.2012.15797},
archivePrefix = {arXiv},
       eprint = {2012.15797},
 primaryClass = {astro-ph.IM},
       adsurl = {https://ui.adsabs.harvard.edu/abs/2020arXiv201215797L}
}

@BOOK{Sersic:1968,
       author = {{Sersic}, Jose Luis},
        title = "{Atlas de Galaxias Australes}",
         year = 1968,
       adsurl = {https://ui.adsabs.harvard.edu/abs/1968adga.book.....S}
}

@article{Phan:2019,
  title={Composable Effects for Flexible and Accelerated Probabilistic Programming in NumPyro},
  author={Phan, Du and Pradhan, Neeraj and Jankowiak, Martin},
  journal={arXiv preprint arXiv:1912.11554},
  year={2019}
}

@ARTICLE{Mowla:2019,
       author = {{Mowla}, Lamiya A. and {van Dokkum}, Pieter and {Brammer}, Gabriel B. and {Momcheva}, Ivelina and {van der Wel}, Arjen and {Whitaker}, Katherine and {Nelson}, Erica and {Bezanson}, Rachel and {Muzzin}, Adam and {Franx}, Marijn and {MacKenty}, John and {Leja}, Joel and {Kriek}, Mariska and {Marchesini}, Danilo},
        title = "{COSMOS-DASH: The Evolution of the Galaxy Size-Mass Relation since z {\ensuremath{\sim}} 3 from New Wide-field WFC3 Imaging Combined with CANDELS/3D-HST}",
      journal = {Astrophysical Journal},
         year = 2019,
        month = jul,
       volume = {880},
       number = {1},
          eid = {57},
        pages = {57},
          doi = {10.3847/1538-4357/ab290a},
archivePrefix = {arXiv},
       eprint = {1808.04379},
 primaryClass = {astro-ph.GA},
       adsurl = {https://ui.adsabs.harvard.edu/abs/2019ApJ...880...57M}
}

@ARTICLE{Lange:2015,
       author = {{Lange}, Rebecca and {Driver}, Simon P. and {Robotham}, Aaron S.~G. and {Kelvin}, Lee S. and {Graham}, Alister W. and {Alpaslan}, Mehmet and {Andrews}, Stephen K. and {Baldry}, Ivan K. and {Bamford}, Steven and {Bland-Hawthorn}, Joss and {Brough}, Sarah and {Cluver}, Michelle E. and {Conselice}, Christopher J. and {Davies}, Luke J.~M. and {Haeussler}, Boris and {Konstantopoulos}, Iraklis S. and {Loveday}, Jon and {Moffett}, Amanda J. and {Norberg}, Peder and {Phillipps}, Steven and {Taylor}, Edward N. and {L{\'o}pez-S{\'a}nchez}, {\'A}ngel R. and {Wilkins}, Stephen M.},
        title = "{Galaxy And Mass Assembly (GAMA): mass-size relations of z < 0.1 galaxies subdivided by S{\'e}rsic index, colour and morphology}",
      journal = {Monthly Notices of the RAS},
         year = 2015,
        month = mar,
       volume = {447},
       number = {3},
        pages = {2603-2630},
          doi = {10.1093/mnras/stu2467},
archivePrefix = {arXiv},
       eprint = {1411.6355},
 primaryClass = {astro-ph.GA},
       adsurl = {https://ui.adsabs.harvard.edu/abs/2015MNRAS.447.2603L}
}

@ARTICLE{Kawinwanichakij:2021,
       author = {{Kawinwanichakij}, Lalitwadee and {Silverman}, John D. and {Ding}, Xuheng and {George}, Angelo and {Damjanov}, Ivana and {Sawicki}, Marcin and {Tanaka}, Masayuki and {Taranu}, Dan S. and {Birrer}, Simon and {Huang}, Song and {Li}, Junyao and {Onodera}, Masato and {Shibuya}, Takatoshi and {Yasuda}, Naoki},
        title = "{Hyper Suprime-Cam Subaru Strategic Program: A Mass-dependent Slope of the Galaxy Size-Mass Relation at z < 1}",
      journal = {Astrophysical Journal},
         year = 2021,
        month = nov,
       volume = {921},
       number = {1},
          eid = {38},
        pages = {38},
          doi = {10.3847/1538-4357/ac1f21},
archivePrefix = {arXiv},
       eprint = {2109.09766},
 primaryClass = {astro-ph.GA},
       adsurl = {https://ui.adsabs.harvard.edu/abs/2021ApJ...921...38K}
}

@ARTICLE{Erwin:2015,
       author = {{Erwin}, Peter},
        title = "{IMFIT: A Fast, Flexible New Program for Astronomical Image Fitting}",
      journal = {Astrophysical Journal},
         year = 2015,
        month = feb,
       volume = {799},
       number = {2},
          eid = {226},
        pages = {226},
          doi = {10.1088/0004-637X/799/2/226},
archivePrefix = {arXiv},
       eprint = {1408.1097},
 primaryClass = {astro-ph.IM},
       adsurl = {https://ui.adsabs.harvard.edu/abs/2015ApJ...799..226E}
}

@ARTICLE{Peng:2002,
       author = {{Peng}, Chien Y. and {Ho}, Luis C. and {Impey}, Chris D. and {Rix}, Hans-Walter},
        title = "{Detailed Structural Decomposition of Galaxy Images}",
      journal = {Astronomical Journal},
         year = 2002,
        month = jul,
       volume = {124},
       number = {1},
        pages = {266-293},
          doi = {10.1086/340952},
archivePrefix = {arXiv},
       eprint = {astro-ph/0204182},
 primaryClass = {astro-ph},
       adsurl = {https://ui.adsabs.harvard.edu/abs/2002AJ....124..266P}
}

@ARTICLE{Robotham:2017,
       author = {{Robotham}, A.~S.~G. and {Taranu}, D.~S. and {Tobar}, R. and {Moffett}, A. and {Driver}, S.~P.},
        title = "{PROFIT: Bayesian profile fitting of galaxy images}",
      journal = {Monthly Notices of the RAS},
         year = 2017,
        month = apr,
       volume = {466},
       number = {2},
        pages = {1513-1541},
          doi = {10.1093/mnras/stw3039},
archivePrefix = {arXiv},
       eprint = {1611.08586},
 primaryClass = {astro-ph.IM},
       adsurl = {https://ui.adsabs.harvard.edu/abs/2017MNRAS.466.1513R}
}

@ARTICLE{Shajib:2019,
       author = {{Shajib}, Anowar J.},
        title = "{Unified lensing and kinematic analysis for any elliptical mass profile}",
      journal = {Monthly Notices of the RAS},
         year = 2019,
        month = sep,
       volume = {488},
       number = {1},
        pages = {1387-1400},
          doi = {10.1093/mnras/stz1796},
archivePrefix = {arXiv},
       eprint = {1906.08263},
 primaryClass = {astro-ph.CO},
       adsurl = {https://ui.adsabs.harvard.edu/abs/2019MNRAS.488.1387S}
}

@article{Hoffman:2014,
  title="{The No-U-Turn sampler: adaptively setting path lengths in Hamiltonian Monte Carlo}",
  author={Hoffman, Matthew D and Gelman, Andrew and others},
  journal={J. Mach. Learn. Res.},
  volume={15},
  number={1},
  pages={1593--1623},
  year={2014}
}

@inproceedings{DeCao:2020,
  title={Block neural autoregressive flow},
  author={De Cao, Nicola and Aziz, Wilker and Titov, Ivan},
  booktitle={Uncertainty in artificial intelligence},
  pages={1263--1273},
  year={2020},
  organization={PMLR}
}

\end{document}